%% file: main.tex
\documentclass[twocolumn]{aastex63}

\usepackage{amsmath,amssymb}
\usepackage{bm}
\usepackage{mathtools}
\usepackage{algorithmic}
\usepackage{algorithm}
\usepackage{textcomp}
\usepackage[nooneline]{subfigure}
\subfiguretopcaptrue
\newcommand{\argmin}{\mathop{\rm argmin}\limits}

\newcommand{\minimize}{\mathop{\rm minimize\ }\limits}
\newcommand{\subjectto}{\mathop{\rm \  subject\, to\ }\limits}

\DeclareMathOperator{\prox}{prox}

\DeclareMathOperator{\diag}{diag}

\usepackage{amsthm}

\usepackage{color}
\usepackage{ulem}

\DeclareRobustCommand{\erase}{\bgroup\markoverwith{\textcolor{red}{\rule[.5ex]{2pt}{0.4pt}}}\ULon} 

\usepackage{comment}
\definecolor{red}{rgb}{0.8,0.0,0.0}
\definecolor{blue}{rgb}{0.0,0.0,0.8}
\definecolor{green}{rgb}{0.0,0.5,0.0}

\received{June 17, 2026}
\revised{September 1, 2026}
\accepted{September 8, 2026}
\submitjournal{ApJ}

\shorttitle{Hybrid Spin--Orbit Tomography for Earth-like Planets}
\shortauthors{Kuwata \& Kawahara}

\begin{document}

\title{Hybrid Spin--Orbit Tomography for Earth-like Planets: Simultaneous Mapping of Static Surfaces and Dynamic Clouds from Multicolor Light Curves}

\correspondingauthor{Atsuki Kuwata}
\email{a.kuwata@astron.s.u-tokyo.ac.jp}

\author[0000-0002-3244-7136]{Atsuki Kuwata}
\affiliation{Department of Astronomy, The University of Tokyo, 7-3-1, Hongo, Bunkyo-ku, Tokyo 113-0033, Japan}
\affiliation{Institute of Space and Astronautical Science, Japan Aerospace Exploration Agency, 3-1-1 Yoshinodai, Chuo-ku, Sagamihara, Kanagawa 252-5210, Japan}

\author[0000-0003-3309-9134]{Hajime Kawahara}
\affiliation{Department of Astronomy, The University of Tokyo, 7-3-1, Hongo, Bunkyo-ku, Tokyo 113-0033, Japan}
\affiliation{Institute of Space and Astronautical Science, Japan Aerospace Exploration Agency, 3-1-1 Yoshinodai, Chuo-ku, Sagamihara, Kanagawa 252-5210, Japan}


\begin{abstract}

Photometric variability of directly imaged exoplanets encodes information on both persistent surface features and time-variable phenomena such as clouds. We present Hybrid Spin--Orbit Tomography, a method that simultaneously retrieves multiple static components and a single dynamic component, together with their reflection spectra, from multiband photometric variability. The method integrates spectral unmixing and dynamic spin--orbit tomography by introducing a time-dependent component into the forward model. We employ sparse regularization for static surface distributions, volume regularization for reflection spectra, and a Kronecker-sum kernel regularization for the dynamic component. Applying the method to a toy Earth model, we successfully recover static surface distributions together with a single time-dependent cloud component. Furthermore, we apply the method to DSCOVR/EPIC observations of Earth and retrieve a dynamic component consistent with real cloud distributions, together with components broadly interpretable as oceans, vegetation, and land surfaces. Relative to static spin--orbit unmixing, the hybrid model substantially reduces the residual scatter of the multiband light curves, demonstrating that explicit treatment of dynamic components improves global mapping from color variability. These results provide a step toward the simultaneous retrieval of static and dynamic planetary components, while highlighting the need for more physical cloud--surface masking models and uncertainty quantification.

\end{abstract}


\input{revised_section_introduction}
\input{revised_section_formulation}
\input{revised_section_optimization}
\input{revised_section_test_toymap}
\input{revised_section_test_DSCOVR}
\input{revised_section_discussion}

\input{revised_section_conclusion}

\acknowledgments

\input{section_acknowledgments}

\appendix
\input{revised_section_kronecker_sum}
\input{revised_section_appendix_proximal}
\input{revised_section_evaluate}
\input{section_appendix_geometry}
\input{section_appendix_residual}

\bibliography{bibtex}{}
\bibliographystyle{aasjournal}

\end{document}

%% file: revised_section_introduction.tex
\section{Introduction}
Following the recommendation of the Astro2020 Decadal Survey, NASA is developing the Habitable Worlds Observatory (HWO), whose primary science goals include the direct imaging and characterization of potentially habitable terrestrial exoplanets \citep{NAP26141,2026arXiv260111803F}.
Alongside reflected-light spectroscopy, time-resolved multiband photometry will provide a complementary probe of planetary rotation and spatially heterogeneous surface properties, including oceans and potential surface biosignatures \citep{2026ASPC..542..439L,2026ASPC..542..393P}.

Such photometric variability provides invaluable information on planetary environments and surface conditions \citep[e.g.][]{ford2001characterization, 2009ApJ...700..915C, 2011ApJ...738..184F}.
The reflected light from a planet is modulated by both its rotation and orbital motion, and these modulations encode two-dimensional information about the planetary surface.
Based on this concept, \citet{kawahara2010global} proposed a two-dimensional inversion technique termed \textit{spin--orbit tomography} (SOT), which reconstructs planetary maps from photometric variations of the planet.
Since then, SOT has been studied in terms of regularization of geography \citep{kawahara2011mapping, fujii2012mapping}, sparse modeling \citep{aizawa2020global, kuwata2022global}, Bayesian formulations \citep{2018AJ....156..146F, kawahara2020bayesian}, non-Lambertian effects \citep{2021arXiv210306275L}, and applications to real Earth data \citep{2019ApJ...882L...1F,2021arXiv210306275L}.

Multiband photometric variability also contains information on the reflection spectra of individual surface components such as water, soil, vegetation, snow, and clouds.
\textit{Rotational spectral unmixing}, which is a method for decomposing light curves into individual surface spectra and their spatial distributions, has been explored using EPOXI observations of the Earth \citep{2013ApJ...765L..17C,2018AJ....156..301L}.
However, the inferred spectra and surface distributions suffer from strong degeneracies because matrix factorization alone generally does not provide a unique solution \citep{2017AJ....154..189F}.
To address this issue, \citet{kawahara2020global} proposed \textit{spin--orbit unmixing} (SOU), which combines SOT and spectral unmixing into a single inverse problem using nonnegative matrix factorization and simplex volume minimization.

Furthermore, \citet{kuwata2022global} introduced sparse modeling into SOU and demonstrated that sparse regularization improves both the inferred maps and the unmixed spectra.
Previous studies on SOT and SOU assumed that surface components remain static during the observation period.
However, in practice, dynamic components such as clouds are likely to exist on the planet and affect the light curve.
To account for such temporal variability, \citet{kawahara2020bayesian} developed a Bayesian framework for dynamic spin--orbit tomography using Gaussian-process priors.
This method successfully retrieved dynamic maps, but focused on single-band or principal-component maps and did not explicitly disentangle spectral and spatial information.

While SOU enables the simultaneous retrieval of surface spectra and geography from multiband photometric variability, it assumes that all planetary components remain static during the observation period.
Conversely, dynamic SOT can reconstruct time-varying surface distributions, but does not explicitly separate spatial and spectral information.
In this study, we combine the spectral extension of SOU with the temporal extension of dynamic SOT.
We explicitly incorporate time-dependent clouds into the forward model and simultaneously infer the dynamic cloud distribution, static surface geography, and reflection spectra from multiband light curves.

The remainder of this paper is organized as follows. In Section~\ref{sec:formulation}, we formulate the simultaneous retrieval of static and dynamic planetary components. In Section~\ref{sec:solve_optimization_problem}, we derive the optimization problem and describe the regularization terms adopted in this study. In Section~\ref{sec:test_toymap}, we validate the proposed method using a toy Earth model containing a time-varying cloud component. In Section~\ref{sec:test_dscovr}, we demonstrate the method by applying it to Earth observations obtained by the Deep Space Climate Observatory (DSCOVR)/Earth Polychromatic Imaging Camera (EPIC). In Section~\ref{sec:discussion}, we discuss the limitations and future directions of the proposed method. In Section~\ref{sec:conclusion}, we summarize our findings.

%% file: revised_section_formulation.tex
\section{Formulation} \label{sec:formulation}

In this section, we first review SOT as a previously proposed method for static mapping, SOU as a static unmixing method that applies SOT to spectral unmixing, and dynamic SOT as an extension of SOT for inferring dynamic components. We then formulate \textit{hybrid} SOT, the spectral unmixing method proposed in this paper, that incorporates dynamic components. Table~\ref{tab:mapping technique} summarizes the range of applicability of the methods introduced in this section.

\begin{table*}[]
    \centering
    \begin{tabular}{lcccl}
        \hline \hline
        Method & Static map & Dynamic map & Spectrum & References\\
        \hline
        spin--orbit tomography (SOT) & \checkmark & - & - & \cite{kawahara2010global} \\
        rotational spectral unmixing & - & - & \checkmark & \cite{2013ApJ...765L..17C}\\
        spin--orbit unmixing (SOU) & \checkmark & - & \checkmark & \cite{kawahara2020global, kuwata2022global} \\
        dynamic SOT & - & \checkmark & \checkmark (PC1) & \cite{kawahara2020bayesian} \\
        hybrid SOT & \checkmark & \checkmark (clouds) & \checkmark & this paper \\
        \hline
    \end{tabular}
    \caption{Scope of applicability of each method and relevant references.}
    \label{tab:mapping technique}
\end{table*}

\subsection{Static Spin--Orbit Tomography}\label{sec:SOT} 

We model the reflected flux from a planetary surface. The reflected planetary flux can be approximated as:
\begin{align}
    \begin{split}
        f_\mathrm{p}^\mathrm{ref} =\frac{f_\star R_\mathrm{p}^2}{\pi a^2}\int_{S_\mathrm{I}\cap S_\mathrm{V}}\mathrm{d}S R(t, \vartheta_0,\varphi_0,\vartheta_1,\varphi_1) \cdot \hspace{8mm} \\
        (\bm{e}_\mathrm{S}\cdot\bm{e}_\mathrm{R}) (\bm{e}_\mathrm{O}\cdot\bm{e}_\mathrm{R}),
    \end{split}
\end{align}
where $R_\mathrm{p}$ denotes the planetary radius; $a$ is the star--planet separation; $f_\star$ is the stellar flux; $R(t, \vartheta_0,\varphi_0,\vartheta_1,\varphi_1)$ is the bidirectional reflectance distribution function (BRDF); $t$ is time; $\vartheta_0$ and $\varphi_0$ are the incident zenith angle and azimuth angle, respectively; $\vartheta_1$ and $\varphi_1$ are the reflected zenith angle and azimuth angle, respectively; $\bm{e}_\mathrm{O}$, $\bm{e}_\mathrm{S}$, and $\bm{e}_\mathrm{R}$ denote the unit vectors from the center of the planet to the observer, primary star, and surface, respectively; and $S_\mathrm{I}\cap S_\mathrm{V}$ denotes the intersection of the illuminated area $S_\mathrm{I}$ ($\bm{e}_\mathrm{S}\cdot\bm{e}_\mathrm{R}>0$) and the visible area $S_\mathrm{V}$ ($\bm{e}_\mathrm{O}\cdot\bm{e}_\mathrm{R}>0$).

Assuming isotropic reflection, the BRDF is expressed as a function of time $t$ and the spherical coordinate $\Omega$:
\begin{align}
    m(t,\Omega)\coloneqq R(t, \vartheta_0,\varphi_0,\vartheta_1,\varphi_1),
\end{align}
where $m(t,\Omega)$ is the albedo of the planetary surface. Then, we obtain
\begin{align}
    f_\mathrm{p}^\mathrm{ref}(t) &= \int \mathrm{d}\Omega G(t,\Omega) m(t,\Omega),
\end{align}
where 
\begin{align}
    G(t,\Omega) &\coloneqq 
    \begin{cases}
    {\displaystyle \frac{f_\star R_\mathrm{p}^2}{\pi a^2} } (\bm{e}_\mathrm{S}\cdot\bm{e}_\mathrm{R}) (\bm{e}_\mathrm{O}\cdot\bm{e}_\mathrm{R}) \\
    \hspace{2.5em} (\bm{e}_\mathrm{S}\cdot\bm{e}_\mathrm{R}>0,\ \bm{e}_\mathrm{O}\cdot\bm{e}_\mathrm{R}>0) \\
    0 \hspace{2em} (\mathrm{otherwise}).
    \end{cases}
\end{align}
Discretizing of $t$ and $\Omega$ yields
\begin{align}
    f_\mathrm{p}^\mathrm{ref}(t_i) &= \sum_j \Delta\Omega G(t_i,\Omega_j) m(t_i,\Omega_j). \label{eq:ref_flux}
\end{align}
Assuming that the surface albedo is static (time-independent), that is, $m(\Omega)\coloneqq m(t,\Omega)$, Equation~\eqref{eq:ref_flux} is written in matrix form as:
\begin{align}
    \bm{d}=W\bm{m}, \label{eq:static_sot}
\end{align}
where $d_i\coloneqq f_\mathrm{p}^\mathrm{ref}(t_i)$; $m_j\coloneqq m(\Omega_j)$; $W_{ij}\coloneqq G(t_i,\Omega_j)\Delta\Omega$ $(i=1,\ldots, N_i, j=1,\ldots, N_j)$; and $N_i$ and $N_j$ are the number of observations and surface pixels, respectively.

In practice, observation noise is introduced into the data. We solve a linear inverse problem in which we infer the surface distribution of the planet $\bm{m}$ from the observational data $\bm{d}$, assuming that $W$ is known. This methodology is called (static) spin--orbit tomography \citep[SOT;][]{kawahara2010global, kawahara2011mapping}. Note that $W$ depends on the orbital parameters of the planet, which can be inferred within the framework of SOT \citep{kawahara2010global,2016MNRAS.457..926S,2018AJ....156..146F,kawahara2020bayesian}; however, in this study, we assume that these nonlinear parameters are known.

\subsection{Static Spin--Orbit Unmixing}\label{sec:SOU}
Rotational spectral unmixing, which attempts to infer surface spectral components from multiband photometric variations in reflected light, was proposed by \cite{2013ApJ...765L..17C}. This study represented an attempt to apply spectral unmixing, a technique widely studied in remote sensing, to exoplanets. In the integrated light from a planet, the mixing fractions of different surface components are generally high, making it difficult to infer individual spectral endmembers. Applying an SOT-like method to multiband data, spin--orbit unmixing \citep[SOU;][]{kawahara2020global} addresses this problem by using information not only along the rotational direction but also along the orbital direction, and by solving spectral unmixing and two-dimensional mapping simultaneously.

The observational data at wavelength $\lambda_l$ is
$\hat{\bm{d}}_l=W\hat{\bm{m}}_l$,
where $\hat{\bm{d}}_l\coloneqq\bm{d}(\lambda_l)$; $\hat{\bm{m}}_l\coloneqq\bm{m}(\lambda_l)$ $(l=1,\ldots,N_l)$; and $N_l$ is the number of observed wavelengths. Then, we can expand Equation~\eqref{eq:static_sot} as:
\begin{align}
    D=WM, \label{eq:static_multi_sot}
\end{align}
where $D\in\mathbb{R}^{N_i\times N_l}$ is a data matrix with $\hat{\bm{d}}_l$ in the $l$-th column and $M\in\mathbb{R}^{N_j\times N_l}$ is a model matrix with $\hat{\bm{m}}_l$ in the $l$-th column.

Moreover, we apply spectral unmixing to $M$, which is a method to decompose observed data into the abundance and the spectra of multiple components, namely,
\begin{align}
    M=AX,
\end{align}
where $A\in\mathbb{R}^{N_j\times N_k}$ denotes the abundance; $X\in\mathbb{R}^{N_k\times N_l}$ denotes the spectra; and $N_k$ denotes the number of components.

In this paper, we consider non-negative matrix factorization (NMF) under the assumption of non-negativity. Thus, Equation~\eqref{eq:static_multi_sot} is expressed as:
\begin{align}
    D&=WA X, \label{eq:static_sou}
\end{align}
Moreover, Equation~\eqref{eq:static_sou} can be written as a sum over individual components:
\begin{align}
    D=\sum_k W \hat{\bm{a}}_k \check{\bm{x}}_k^\top, \label{eq:static_sou_k}
\end{align}
where $\hat{\bm{a}}_k$ is the $k$-th column vector of $A$ and $\check{\bm{x}}_k^\top$ is the $k$-th row vector of $X$.
This methodology is called spin--orbit unmixing \citep[SOU;][]{kawahara2020global, kuwata2022global}.

\subsection{Dynamic Spin--Orbit Tomography}

We also review dynamic SOT, a technique that extends SOT to handle time-varying geography. 
Equation~\eqref{eq:ref_flux} can be rewritten in a time-dependent form as:
\begin{align}
    d_i = \sum_j W_{ij} m^{(i)}_j \ \ \ \ \left(i=1,\ldots,N_i\right),
\end{align}
where $m^{(i)}_j\coloneqq m(t_i,\Omega_j)$.

Here, we define the dynamic geography matrix $C \in \mathbb{R}^{N_i\times N_j}$ as $C_{ij} \coloneqq m^{(i)}_j$; thus we obtain:
\begin{align}
    \bm{d} =\left( W \odot C \right)\bm{1}_{N_j}, \label{eq:dynamic_sot_Hadamard}
\end{align}
where $\bm{1}_N$ denotes the vector of ones with dimension $N$; and $\odot$ denotes the Hadamard product.
\citet{kawahara2020bayesian} developed a Bayesian framework for dynamic SOT and successfully reconstructed dynamic maps using the principal component of real observational data.


\subsection{Hybrid Spin--Orbit Tomography}

In this paper, we propose \textit{hybrid spin--orbit tomography}, which integrates static SOU and dynamic SOT. In this framework, we simultaneously estimate $N_k$ static surface distributions, a single dynamic component distribution, and $(N_k+1)$ reflection spectra. Combining \eqref{eq:static_sou} and \eqref{eq:dynamic_sot_Hadamard}, the observational data can be expressed as:
\begin{align}
    D = \left( W \odot C \right)\bm{1}_{N_j} \bm{x}_\mathrm{C}^\top + WAX, \label{eq:hybrid_simple_sum}
\end{align}
where $\bm{x}_\mathrm{C} \in \mathbb{R}^{N_l}$ denotes the reflection spectrum of the dynamic component.

%% file: revised_section_optimization.tex
\section{Optimization Problem} \label{sec:solve_optimization_problem}

Our methodology seeks to estimate $C$, $\bm{x}_\mathrm{C}$, $A$, and $X$ given the observed data $D$ and the known design matrix $W$. This is a type of inverse problem, and we can express it as a mathematical optimization problem:
\begin{align}
    & \minimize_{C, \bm{x}_\mathrm{C}, A, X} F , \\
    &F \coloneqq \left\|D - \left( W \odot C \right)\bm{1}_{N_j} \bm{x}_C^\top - WAX \right\|_\mathrm{F}^2,
\end{align}
where $\|\cdot\|_\mathrm{F}$ denotes the Frobenius norm.
In practice, we add regularization terms $R$ and constraints $P$ to deal with model instability due to observational noise and non-uniqueness arising from NMF:
\begin{align}
    & \minimize_{C, \bm{x}_\mathrm{C}, A, X} Q  \subjectto P \left( C, \bm{x}_\mathrm{C}, A, X \right) \label{eq:optimization_problem_general} \\
    & Q \coloneqq F \left( C, \bm{x}_\mathrm{C}, A, X \right) + R \left( C, \bm{x}_\mathrm{C}, A, X \right).
\end{align}

\subsection{Reflection Spectra and Static Surface Components}

First, the regularization terms and constraints for $\bm{x}_\mathrm{C}, A,$ and $X$ are:
\begin{align}
    A \ge O, X \ge O, \bm{x}_\mathrm{C} \ge \bm{0}, \label{eq:constraints_A_X_xC}
\end{align}
\begin{align}
    R_A(A) &\coloneqq \sum_{k=1}^{N_k} \left(\lambda_{\ell_1}\| \hat{\bm{a}}_k \|_1 + \lambda_{\mathrm{TSV}}\|\hat{\bm{a}}_k\|_{\mathrm{TSV}}\right), \label{eq:regularization_A} \\ 
    R_{\tilde{X}}\left(\tilde{X}\right) &\coloneqq \lambda_{\tilde{X}} \det \left(\tilde{X}\tilde{X}^\top\right), \label{eq:regularization_X} \\ 
    \tilde{X} &\coloneqq \left(
    \begin{array}{cc}
         \bm{x}_\mathrm{C}^\top \\
         X
    \end{array}
    \right) \in \mathbb{R}^{\left(N_k+1\right) \times N_l},
\end{align}
where $A \ge O, X \ge O,$ and $\bm{x}_\mathrm{C} \ge \bm{0}$ denote that all components of $A$, $X$, and $\bm{x}_\mathrm{C}$ are non-negative; $\lambda_{\ell_1}$, $\lambda_{\mathrm{TSV}}$, and $\lambda_{\tilde{X}}$ denote regularization parameters; and $\|\cdot\|_1$ and $\|\cdot\|_\mathrm{TSV}$ denote the $\ell_1$-norm and Total Squared Variation (TSV) norm, respectively. These norms are defined as follows \citep[see Appendix~C in][]{kuwata2022global}:
\begin{align}
    \|\bm{a}\|_1 &\coloneqq \sum_{s} |a_s|, \\
    \|\bm{a}\|_\mathrm{TSV} &\coloneqq \sum_{s,s'} \frac{1}{2} N_{ss'} (a_s-a_{s'})^2, \\
    N_{ss'} &\coloneqq 
        \begin{cases}
            1 &
            \left(
            \begin{aligned}
                &\mathrm{if}\ s\mathrm{th\ and}\ s'\mathrm{th\ pixels} \\
                &\mathrm{are\ adjacent}
            \end{aligned}
            \right) \\
            0 & \left( \mathrm{otherwise} \right).
        \end{cases}
\end{align}
The $\ell_1$-norm regularization promotes sparsity in the solution, meaning that most of the elements are zero.
TSV regularization promotes smoother boundaries between adjacent pixels and reduces noise in the solution.
Furthermore, volume regularization $\det \left(\tilde{X}\tilde{X}^\top\right)$ is effective for resolving spectral components in spectral unmixing, which was developed in the field of remote sensing \citep{craig1994minimum,fu2015blind,fu2019nonnegative,lin2015identifiability,ang2019algorithms}. These regularization terms are also used in static SOU \citep{kuwata2022global}, and their effectiveness has been demonstrated.

\subsection{Dynamic Components}

Next, we consider the regularization terms and constraints for $C$. The constraints for $C$ are:
\begin{align}
    C \ge O, \label{eq:constraints_C}
\end{align}
same as for the other components.

\subsubsection{Kronecker-Product kernel}
Referring to \citet{kawahara2020bayesian}, we consider the following regularization term for $C$:
\begin{align}
    & \mathrm{vec}(C)^\top \left[ \alpha \left(K_{\mathrm{S}} \otimes K_{\mathrm{T}} \right)\right]^{-1} \mathrm{vec}(C) \\
    &= \alpha^{-1}\mathrm{vec}(C)^\top \left(K_{\mathrm{S}}^{-1} \otimes K_{\mathrm{T}}^{-1} \right) \mathrm{vec}(C), \label{eq:RC_Kronecker_product}
\end{align}
where $K_{\mathrm{S}}\in\mathbb{R}^{N_j \times N_j}$ and $K_{\mathrm{T}}\in\mathbb{R}^{N_i \times N_i}$ denote the spatial and temporal kernels, respectively; $\mathrm{vec}$ denotes the vectorization operator
\footnote{
The $\mathrm{vec}$~operator stacks the columns of a matrix into a vector. For any $B = \left( \hat{\bm{b}}_1 \cdots \hat{\bm{b}}_n \right)\in \mathbb{R}^{m \times n}$, $\mathrm{vec}(B)$ is defined as:
\begin{align}
    \mathrm{vec}(B) \coloneqq
    \left(\begin{array}{c}
        \hat{\bm{b}}_1 \\
        \vdots \\
        \hat{\bm{b}}_n
    \end{array}\right)
    \in \mathbb{R}^{mn}. \nonumber
\end{align}
}
; and $\otimes$ denotes the Kronecker product.
This term is based on the fact that we can interpret $\alpha \left(K_{\mathrm{S}} \otimes K_{\mathrm{T}} \right)$ as the covariance matrix of a prior distribution of $\mathrm{vec}(C) \in \mathbb{R}^{N_i N_j}$, where $\alpha$ denotes its amplitude. Further details on this interpretation are provided in Appendix~\ref{sec:kronecker}.

In general, kernel functions express similarities between variables. Representative examples of kernel functions are the radial basis function (RBF) and the Mat\'{e}rn-3/2 kernel:
\begin{align}
    \kappa_{\mathrm{RBF}}\left(\eta;\gamma\right) &\coloneqq \exp\left(-\frac{\eta^2}{2\gamma^2}\right), \\
    \kappa_{\mathrm{M3/2}}\left(\eta;\gamma\right) &\coloneqq \left(1+\frac{\sqrt{3}\eta}{\gamma}\right) \exp\left(-\frac{\sqrt{3}\eta}{\gamma}\right).
\end{align}
We use the RBF kernel as the spatial kernel and the Mat\'{e}rn-3/2 kernel as the temporal kernel, respectively:
\begin{align}
    \left(K_{\mathrm{S}}\right)_{jj'} &= \kappa_{\mathrm{RBF}}\left(\eta_{jj'};\gamma\right), \\
    \left(K_{\mathrm{T}}\right)_{ii'} &= \kappa_{\mathrm{M3/2}}\left(\left|t_i-t_{i'}\right|;\tau\right),
\end{align}
where $\eta_{jj'}$ denotes the angular separation between the $j$th and $j'$th pixels; $t_i$ denotes the $i$th observation time; and $\gamma$ and $\tau$ denote the spatial and temporal correlation scales, respectively.

Although the Kronecker-product kernel can model spatial and temporal correlations simultaneously, it uses a single amplitude parameter. Therefore, the strengths of the spatial and temporal regularization terms cannot be controlled independently. 

\subsubsection{Kronecker-Sum kernel}
To address this limitation, we consider the Kronecker-sum kernel:
\begin{align}
    &\lambda_\mathrm{S} K_{\mathrm{S}}^{-1} \oplus \lambda_\mathrm{T} K_{\mathrm{T}}^{-1} \\
    &= \lambda_\mathrm{S} K_{\mathrm{S}}^{-1} \otimes I_{N_i} + I_{N_j} \otimes \lambda_\mathrm{T} K_{\mathrm{T}}^{-1},
\end{align}
where the second line follows from the definition of the Kronecker sum; $\lambda_\mathrm{S}$ and $\lambda_\mathrm{T}$ denote regularization parameters.
Using this kernel, we employ the following regularization term for $C$:
\begin{align}
    \mathrm{vec}(C)^\top \left(\lambda_\mathrm{S} K_{\mathrm{S}}^{-1} \oplus \lambda_\mathrm{T} K_{\mathrm{T}}^{-1}\right) \mathrm{vec}(C). \label{eq:RC_kronecker_sum}
\end{align}

In addition to the quadratic kernel, we also consider $\ell_1$-norm regularization for $C$ to promote sparsity in the solution:
\begin{align}
    \lambda_{\ell_1}^{\mathrm{(C)}} \| C \|_1,
\end{align}
where $\lambda_{\ell_1}^{\mathrm{(C)}}$ denotes a regularization parameter. 

Therefore, the regularization term for $C$ is:
\begin{align}
    R_C (C) &=  \lambda_{\ell_1}^{\mathrm{(C)}} \| C \|_1 \nonumber \\
    &+ \mathrm{vec}(C)^\top \left(\lambda_\mathrm{S} K_{\mathrm{S}}^{-1} \oplus \lambda_\mathrm{T} K_{\mathrm{T}}^{-1}\right) \mathrm{vec}(C). \label{eq:regularization_C}
\end{align}


\subsection{Final Form of the Optimization Problem}

Summarizing Equations~\eqref{eq:optimization_problem_general}, \eqref{eq:constraints_A_X_xC}, \eqref{eq:regularization_A}, \eqref{eq:regularization_X}, \eqref{eq:constraints_C}, and \eqref{eq:regularization_C}, the optimization problem to be solved is:
\begin{align}
    \begin{split}
        & \minimize_{C, \bm{x}_\mathrm{C}, A, X} Q \left( C, \bm{x}_\mathrm{C}, A, X \right) \\
        &\subjectto C \ge O, \bm{x}_\mathrm{C} \ge \bm{0}, A \ge O, X \ge O,
    \end{split} \label{eq:optimization_hybrid_sot}
\end{align}
where
\begin{align}
    & Q \left( C, \bm{x}_\mathrm{C}, A, X \right) \nonumber \\
    \begin{split}
        & = F \left( C, \bm{x}_\mathrm{C}, A, X \right) + R_A(A) + R_{\tilde{X}}\left(\tilde{X}\right) + R_C (C)
    \end{split} \\
    \begin{split}
        & = \left\|D - \left( W \odot C \right)\bm{1}_{N_j} \bm{x}_C^\top - WAX \right\|_\mathrm{F}^2 \\
        & \hspace{8mm} + \lambda_{\ell_1}^{\mathrm{(C)}} \| C \|_1 \\
        & \hspace{8mm} + \mathrm{vec}(C)^\top \left(\lambda_\mathrm{S} K_{\mathrm{S}}^{-1} \oplus \lambda_\mathrm{T} K_{\mathrm{T}}^{-1}\right) \mathrm{vec}(C) \\
        & \hspace{8mm} + \sum_{k=1}^{N_k} \left(\lambda_{\ell_1}\|\hat{\bm{a}}_k\|_1 + \lambda_{\mathrm{TSV}}\|\hat{\bm{a}}_k\|_{\mathrm{TSV}}\right) \\
        & \hspace{8mm} + \lambda_{\tilde{X}} \det \left(\tilde{X}\tilde{X}^\top\right).
    \end{split}
\end{align}
Here, we refer to the dynamic component as Component 0 and the static components as Components 1 to $N_k$.

To solve the optimization problem~\eqref{eq:optimization_hybrid_sot}, we use the block coordinate descent method, which divides the problem into subproblems with respect to $C$, $\bm{x}_\mathrm{C}$, $\hat{\bm{a}}_k$, and $\check{\bm{x}}_k$ $(k=1,\ldots, N_k)$, solves the subproblems for each variable, and updates the variables iteratively. Each subproblem is solved using the proximal gradient method (PG). In particular, the PG for $\ell_1$-norm regularization is also called the Iterative Shrinkage-Thresholding Algorithm (ISTA). This method is explained in Appendix~\ref{sec:proximal}.

Static SOU with sparse modeling \citep{kuwata2022global} employed the accelerated PG with the restart method (APG+restart), a type of PG using Nesterov's acceleration \citep{nesterov2003introductory}, to estimate the reflectance spectra.
It also employed the Monotone Fast ISTA \citep[MFISTA;][]{beck2009fastIEEE}, which guarantees monotonic decrease of the objective function, to estimate the geographical distributions.
In this study, we also employ APG+restart and MFISTA to solve the corresponding subproblems.

Consequently, the optimization algorithm for solving \eqref{eq:optimization_hybrid_sot} is summarized as follows:
\begin{algorithm}[H]
    \caption{Hybrid Spin--Orbit Tomography}
    \label{alg:SOU_using_L1TSV}
    \begin{algorithmic}
    \STATE Initialization: non-negative $ C_0$, $ (\bm{x}_\mathrm{C})_0$, $A_0$, $X_0$
    \FOR{$n$ in $(1,N_\mathrm{try})$}
    \STATE Update $\bm{x}_\mathrm{C}$ using APG+restart
     \STATE Update $C$ using MFISTA
    \FOR{$k$ in $(1,N_k)$}
    \STATE Update $\check{\bm{x}}_k$ using APG+restart
    \STATE Update $\hat{\bm{a}}_k$ using MFISTA
    \ENDFOR
    \ENDFOR
    \end{algorithmic}
\end{algorithm}

%% file: revised_section_test_toymap.tex
\section{Experiment Using a Toy Model} \label{sec:test_toymap}


\subsection{Setup of the Toy Earth Model}

\begin{figure*}
  \centering
  \subfigure[]{%
    \includegraphics[width=.45\textwidth]{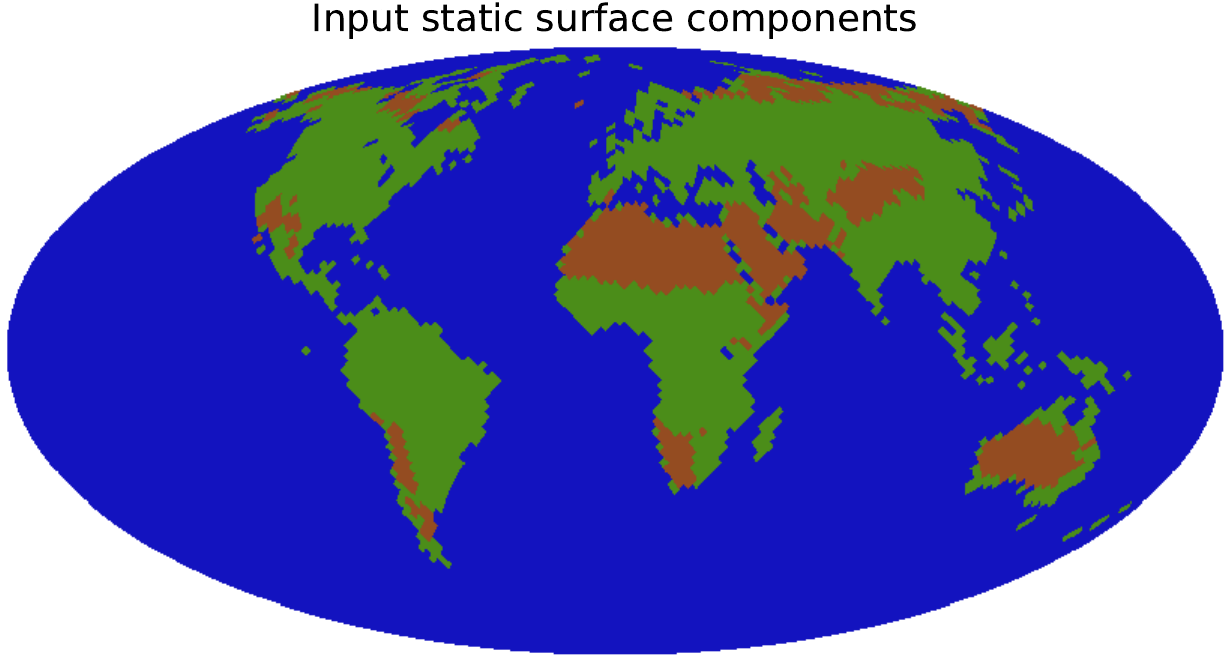}%
    \label{fig:init_surface}%
  }%
  \hspace{.05\textwidth}
  \subfigure[]{%
    \includegraphics[width=.35\textwidth]{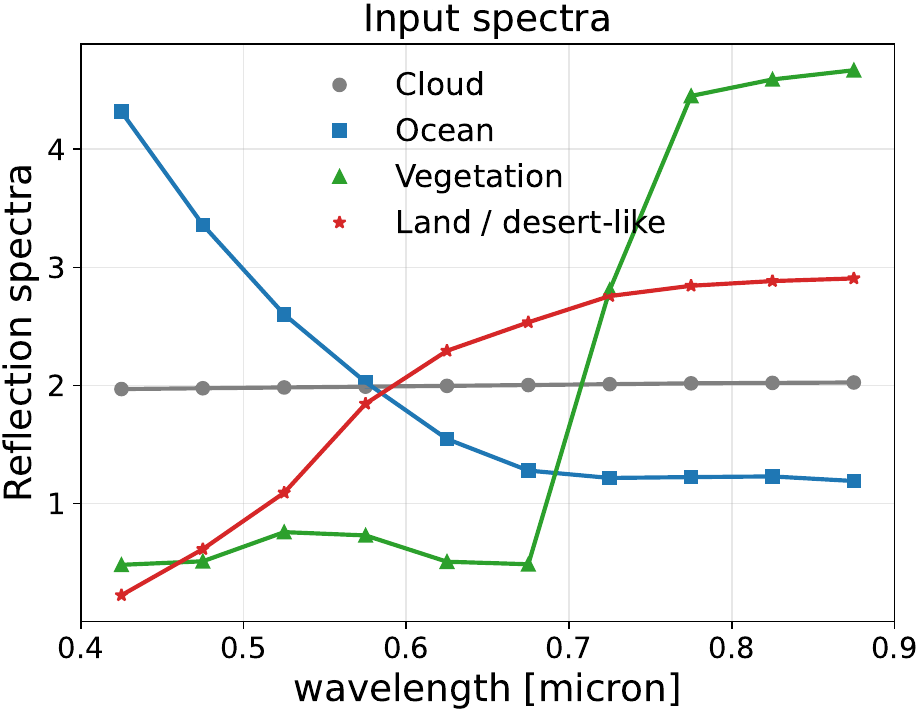}
    \label{fig:init_spectra}%
  }
  \subfigure[]{%
    \includegraphics[width=.99\textwidth]{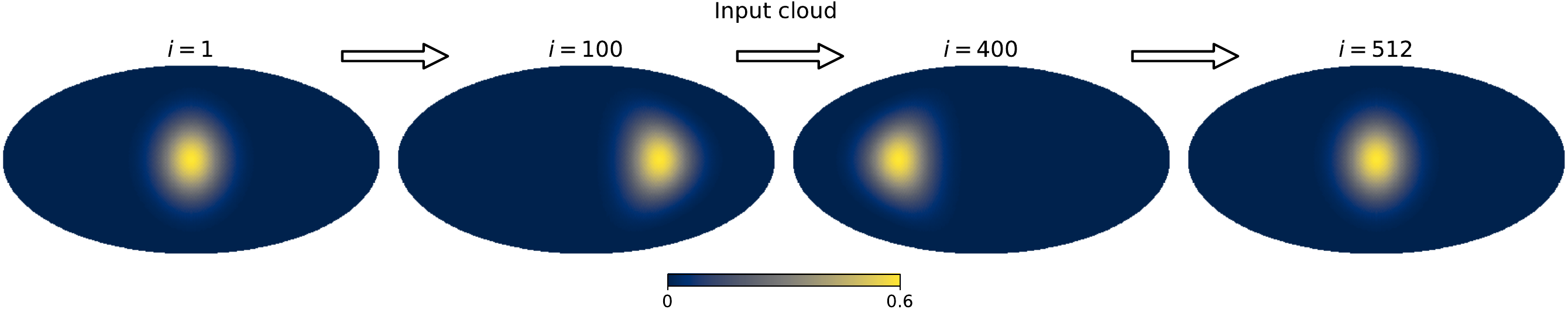}
    \label{fig:init_cloud}%
  }
  \caption[]{Input data for the toy Earth model. (a) Input surface distribution: green is vegetation, brown is land, and blue is ocean. (b) Input spectra of the static surface components and cloud. (c) Input cloud distribution.}
\label{fig:init_surface_spectra}
\end{figure*}

In this section, we test our method using a toy model of the Earth. The static components are the same as in \citet{kuwata2022global}, and the dynamic component is a highly simplified cloud to facilitate verification of the proposed methodology.

We generate mock dynamic multicolor light curves using Equation~\eqref{eq:hybrid_simple_sum} with 1\% observational noise. The observation period is 1~yr and is sampled at $N_i=512$ time points. Furthermore, $N_j^{\mathrm{(true)}}=12{,}288$, $N_k^{\mathrm{(true)}}=3$, and $N_l=10$ denote the number of pixels on the planet surface, (static) surface components, and observing bands, respectively. We considered vegetation, land, and ocean as the end-members of the surface components. We used the classification map provided by the Moderate Resolution Imaging Spectroradiometer (MODIS) as the input surface distribution, the ASTER spectral library \citep{baldridge2009aster} for the vegetation and land spectra, and those produced by \citet{mclinden1997estimating} for ocean spectra. We set the observation wavelengths to $0.425+0.05(l-1)$ \textmu m $ (l=1,\ldots,N_l=10)$. The input surface components and spectra are shown in Figures~\ref{fig:init_surface} and \ref{fig:init_spectra}, respectively.

For the dynamic component, we adopt a toy cloud model consisting of a single Gaussian-like cloud blob that drifts longitudinally with time. The cloud is centered near the equator with an initial latitude and longitude of $0^\circ$ and $0^\circ$, respectively, and moves eastward in longitude. It takes 1~yr to complete a full cycle. The spatial extent is controlled by a Gaussian width of $20^\circ$. We use the cloud spectrum provided by \citet{kokhanovsky2004optical}. The input cloud distribution is shown in Figure~\ref{fig:init_cloud}.

We also constructed the design matrix $W$ with orbital inclination $i=45^\circ$, orbital phase angle at the vernal equinox $\Theta_\mathrm{eq}=90^\circ$, obliquity $\zeta=23^\circ\!\!.4$, orbital period $P_\mathrm{orb}=365.242190402$~days, and rotation period $P_\mathrm{spin}=23.9344699/24.0$~days. Finally, we generated the data matrix $D$ using Equation~\eqref{eq:hybrid_simple_sum}.

\subsection{Estimation}

We solved the optimization problem \eqref{eq:optimization_hybrid_sot}, setting the number of pixels in the estimated map to $N_j=3{,}024$ and the number of surface components to $N_{k}=3$. 
We set the hyperparameters to
$ (\lambda_{\ell_1}^{\mathrm{(C)}}$, $ \lambda_\mathrm{S}$, $ \lambda_\mathrm{T}$, $ \lambda_{\ell_1}$, $ \lambda_\mathrm{TSV}$, $ \lambda_{\tilde{X}}) $ $=$ $ (10^{-2}$, $ 10^{-5}$, $ 10^{-5}$, $ 10^{-3}$, $ 10^{-2}$, $ 10^{0})$. These hyperparameters were selected based on several criteria: e.g., the mean residual, the Correct Pixel Rate (CPR), and the correlation between the estimated and true cloud distributions.
The detailed procedure for hyperparameter selection is presented in Appendix~\ref{sec:evaluate}.
In addition, we selected the spatial correlation scale $\gamma=30^\circ$, which is somewhat wider than the true distribution, and the temporal correlation scale $\tau=1\mathrm{[yr]}$, which is reasonable given that the toy cloud completes one cycle in one year. Note that these hyperparameters should be selected using Bayesian inference; however, this is beyond the scope of this paper.

\begin{figure*}
  \centering
  \subfigure[]{%
    \includegraphics[width=.45\textwidth]{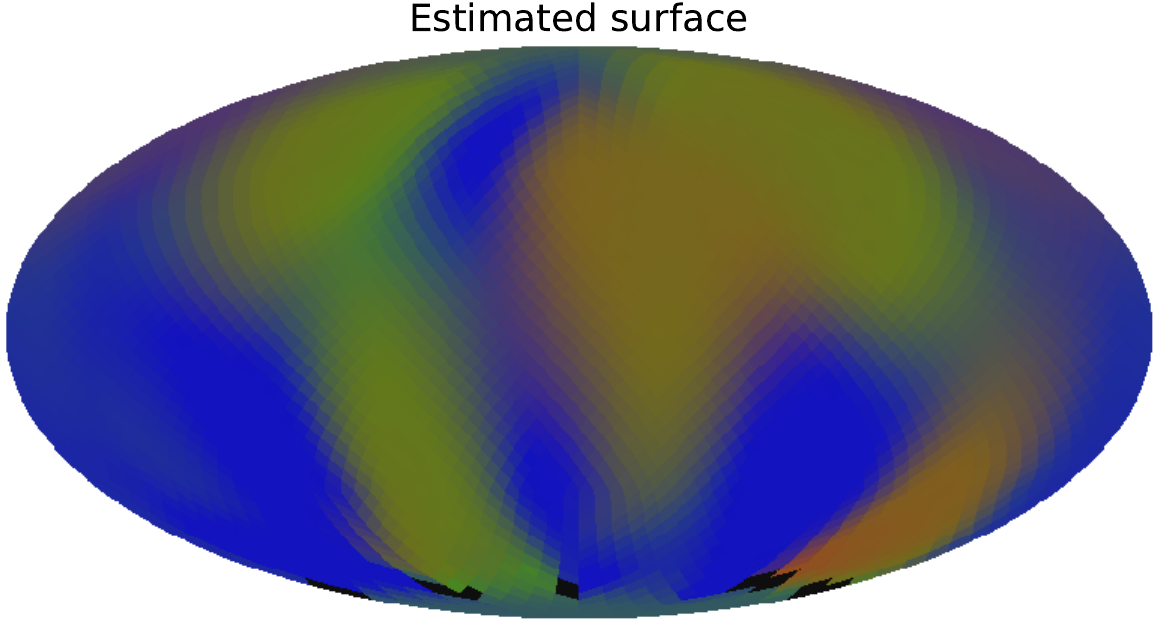}%
    \label{fig:toymap_estimated_surface}%
  }%
  \hspace{.05\textwidth}
  \subfigure[]{%
    \includegraphics[width=.35\textwidth]{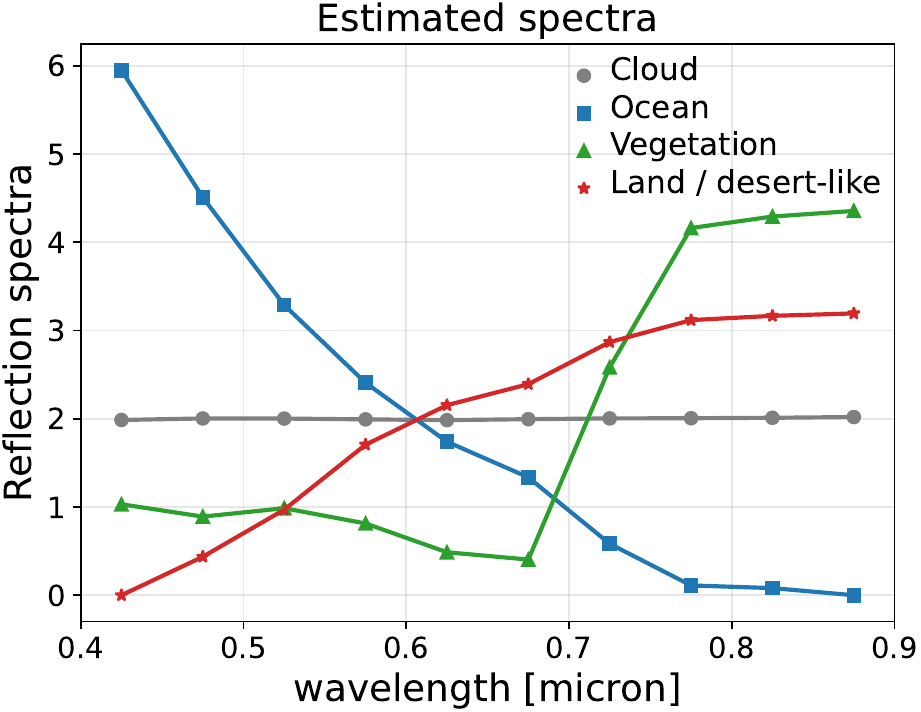}
    \label{fig:toymap_estimated_spectra}%
  }
  \subfigure[]{%
    \includegraphics[width=.99\textwidth]{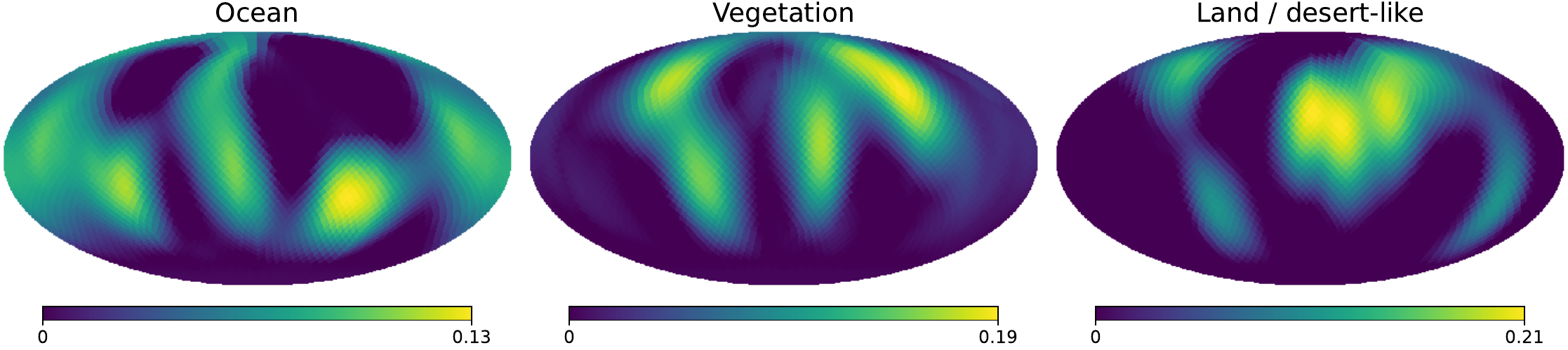}
    \label{fig:toymap_estimated_surface_components}%
  }
  \subfigure[]{%
    \includegraphics[width=.99\textwidth]{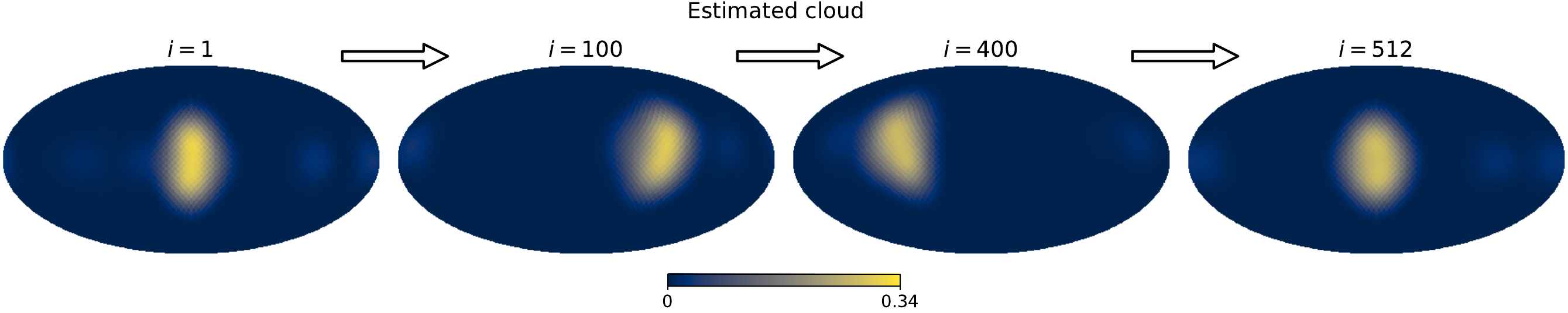}
    \label{fig:toymap_estimated_cloud}%
  }
  \caption[]{The results of the test using the toy model. (a) Color composite of estimated static surface distribution: green is vegetation, brown is land, and blue is ocean. (b) Estimated spectra. (c) Estimated distribution for each static component. (d) Snapshots of the estimated dynamic component (cloud).}
\label{fig:toymap_estimate}
\end{figure*}

Figure~\ref{fig:toymap_estimate} shows the estimated cloud distribution, surface distributions, and spectra. 
For the surface distribution of the static components, shown in Figure~\ref{fig:toymap_estimated_surface} and \ref{fig:toymap_estimated_surface_components}, we successfully captured the continental structure and distinguished vegetation from land in areas such as around the Sahara Desert. 
For the spectra, shown in Figure~\ref{fig:toymap_estimated_spectra}, we succeeded in retrieving the structure of each component. 
The difference in scaling compared to the true spectra is considered to be due to the scale ambiguity of matrix factorization. 
The structure of the dynamic component (cloud) distribution was also captured, as shown in Figure~\ref{fig:toymap_estimated_cloud}, and we were particularly successful in reconstructing temporal variations. 

\begin{figure*}
  \centering
  \subfigure[]{%
    \includegraphics[width=.45\textwidth]{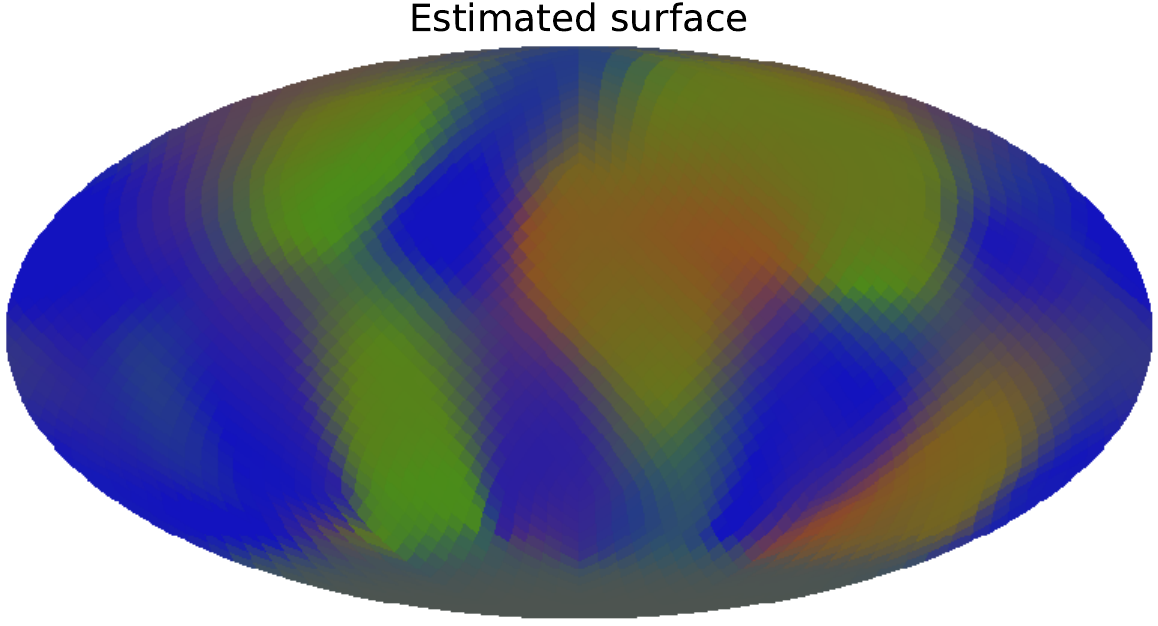}%
    \label{fig:toymap_estimated_surface_v2}%
  }%
  \hspace{.05\textwidth}
  \subfigure[]{%
    \includegraphics[width=.35\textwidth]{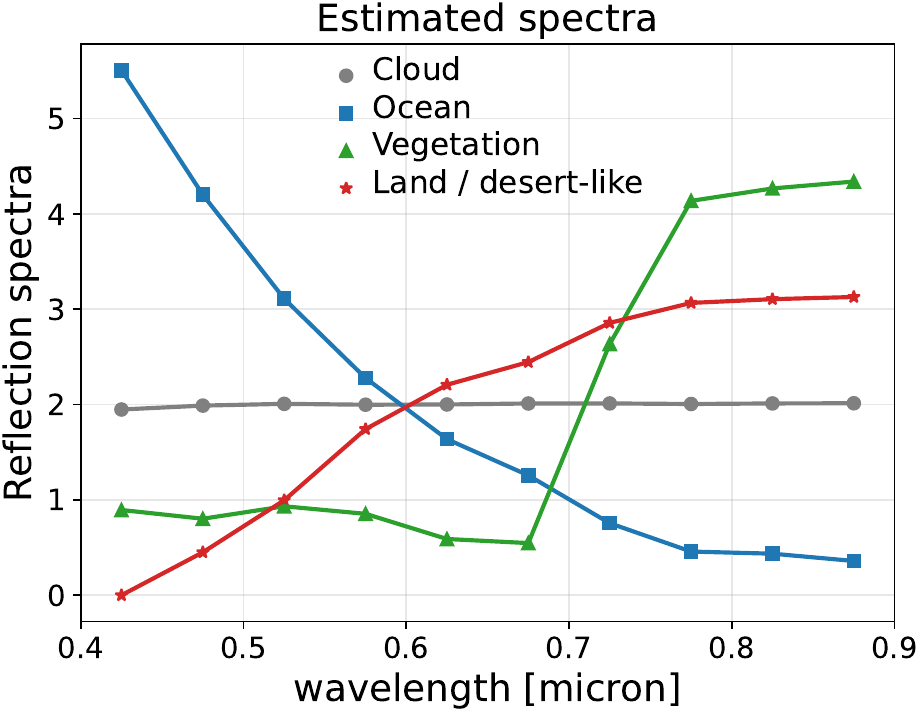}
    \label{fig:toymap_estimated_spectra_v2}%
  }
  \subfigure[]{%
    \includegraphics[width=.99\textwidth]{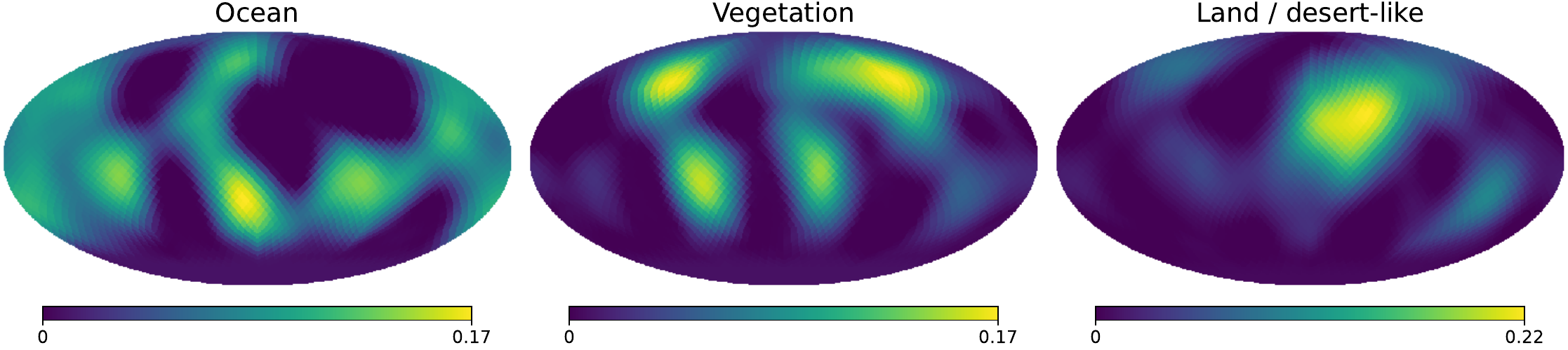}
    \label{fig:toymap_estimated_surface_components_v2}%
  }
  \subfigure[]{%
    \includegraphics[width=.99\textwidth]{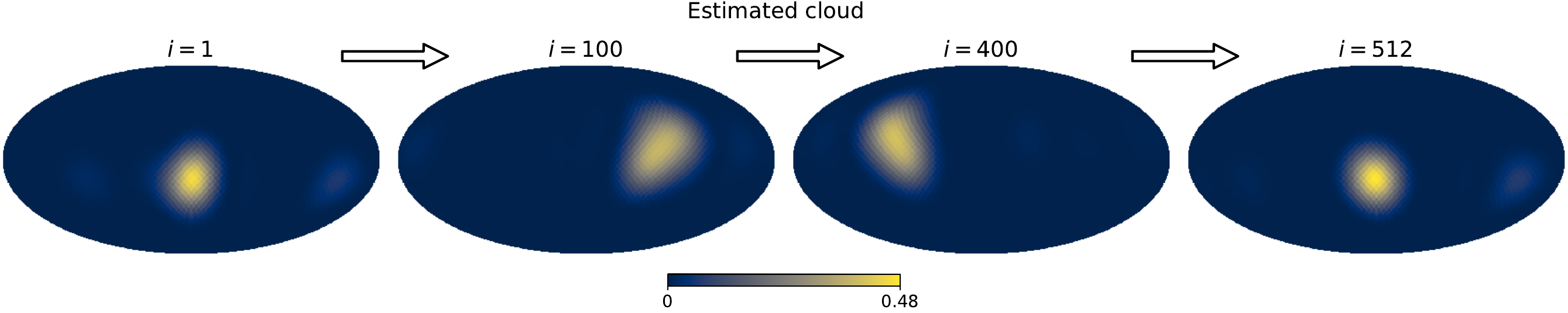}
    \label{fig:toymap_estimated_cloud_v2}%
  }
  \caption[]{Same as Figure~\ref{fig:toymap_estimate}, but for $i=0^\circ$ and $\zeta=60^\circ$.}
\label{fig:toymap_estimate_v2}
\end{figure*}

In addition, we constructed a toy model with a different observing geometry, $i=0^\circ$ and $\zeta=60^\circ$, and applied the proposed method.
Figure~\ref{fig:toymap_estimate_v2} shows the estimated cloud distribution, surface distributions, and spectra under this setting.
We set the hyperparameters to
$ (\lambda_{\ell_1}^{\mathrm{(C)}}$, $ \lambda_\mathrm{S}$, $ \lambda_\mathrm{T}$, $ \lambda_{\ell_1}$, $ \lambda_\mathrm{TSV}$, $ \lambda_{\tilde{X}}) $ $=$ $ (10^{-2.5}$, $ 10^{-5}$, $ 10^{-5}$, $ 10^{-3}$, $ 10^{-1}$, $ 10^{0})$, which were determined based on the same criteria as above.
In this case as well, we successfully retrieved the planetary components.
This provides an initial indication that the method can be applied under different observing geometries.
A more systematic assessment of the sensitivity of the retrieval results to observing geometry is presented in Appendix~\ref{app:geometry}.
Taken together, the proposed hybrid SOT successfully recovered both static and dynamic components simultaneously.

%% file: revised_section_test_DSCOVR.tex
\section{Application to Real Observed Data} \label{sec:test_dscovr}

In this section, we apply our method to real long-term monitoring data of Earth as observed by DSCOVR/Earth Polychromatic Imaging Camera \citep[EPIC;][]{jiang2018using}. DSCOVR has been continuously observing the dayside of Earth from the first Sun--Earth Lagrangian point (L1) since 2015. Although DSCOVR observations are not identical to direct imaging observations of exoplanets, Earth's obliquity together with the spacecraft's orbital motion contributes to the two-dimensional information contained in the observed light curves about the planetary surface. This makes it possible to apply two-dimensional mapping methods \citep{2019ApJ...882L...1F,aizawa2020global,kawahara2020global,kawahara2020bayesian,kuwata2022global}.

\begin{figure*}
  \centering
  \subfigure[]{%
    \includegraphics[width=.45\textwidth]{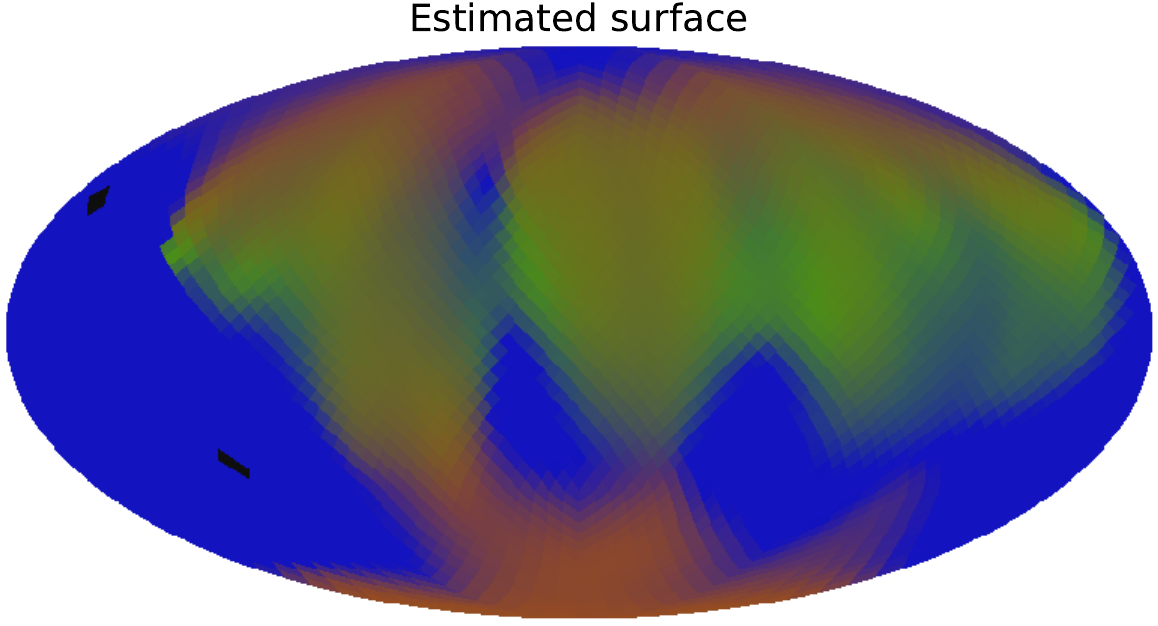}%
    \label{fig:dscovr_estimated_surface}%
  }%
  \hspace{.05\textwidth}
  \subfigure[]{%
    \includegraphics[width=.35\textwidth]{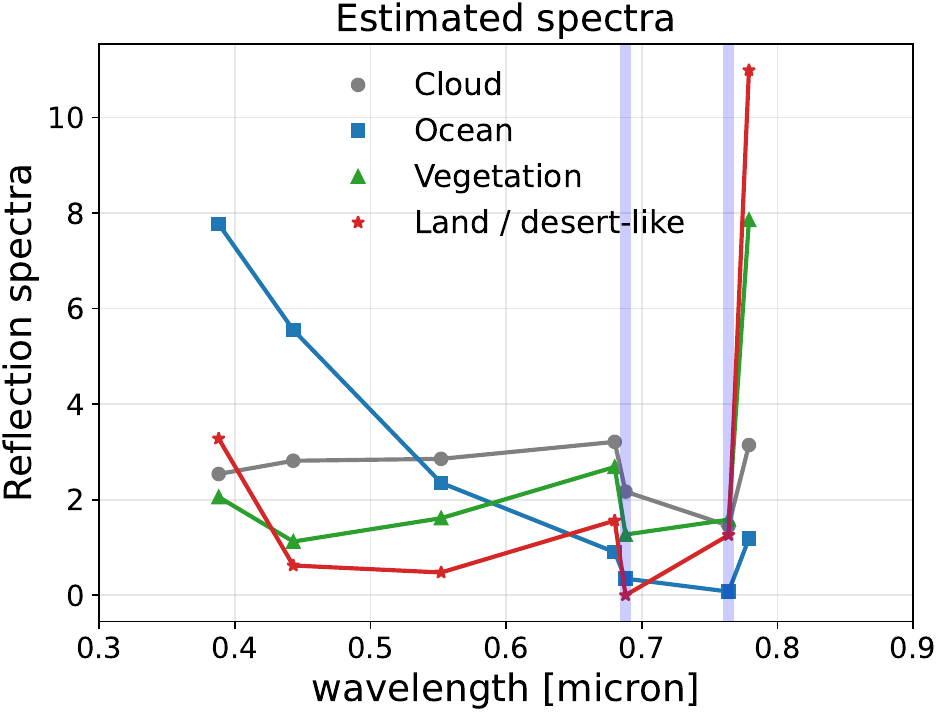}
    \label{fig:dscovr_estimated_spectra}%
  }
  \subfigure[]{%
    \includegraphics[width=.99\textwidth]{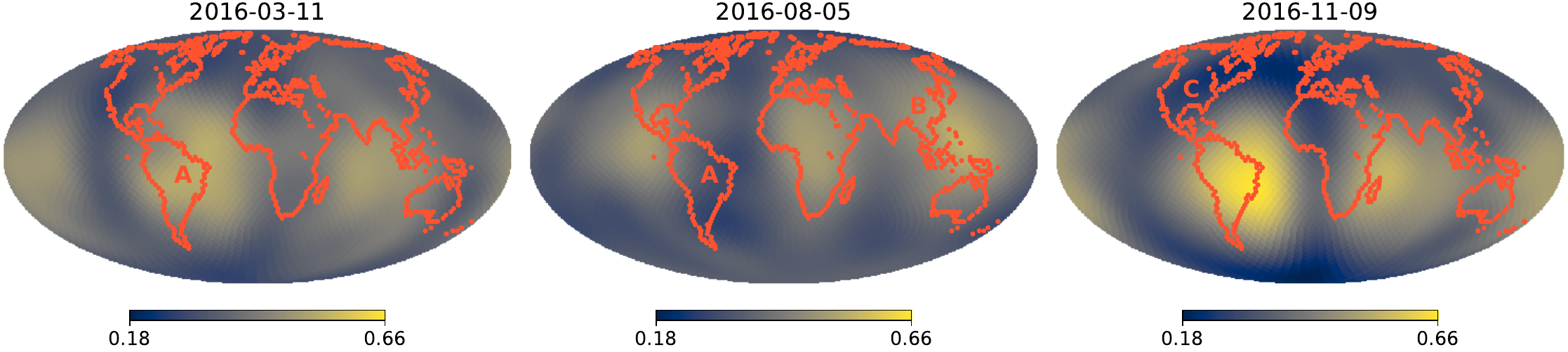}
    \label{fig:dscovr_estimated_cloud}%
  }
  \subfigure[]{%
    \includegraphics[width=.99\textwidth]{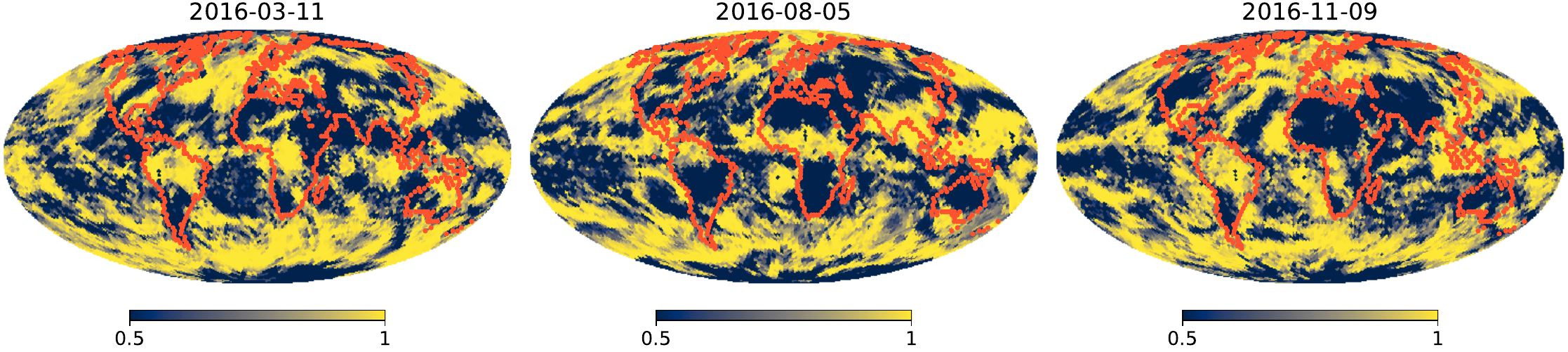}
    \label{fig:dscovr_true_cloud}%
  }
  \caption[]{The results of the test using DSCOVR data. (a) Estimated static surface distribution. (b) Estimated spectra. (c) Snapshots of the estimated dynamic component (cloud) for three different dates in March, August, and November 2016 (from left to right). (d) Observed mean cloud fraction over 8 days.}
\label{fig:dscovr_estimate}
\end{figure*}

Following the same setup as that in \citet{kuwata2022global}, we used one-quarter of the 2~yr data of 2016 and 2017 (i.e., one from each of the four bins) used in \citet{2019ApJ...882L...1F}. As a result, the number of observation frames is $N_i = 2435$. The observed wavelengths are $N_l = 7$ optical bands used in the EPIC instrument (0.388, 0.443, 0.552, 0.680, 0.688, 0.764, and 0.779 \textmu m). There are strong oxygen B and A absorptions at 0.688 and 0.764 \textmu m, respectively. 
Moreover, we set the number of pixels in the estimated map to $N_j=3{,}024$ and the number of surface components to $N_{k}=3$. 
We set the hyperparameters to
$ (\lambda_{\ell_1}^{\mathrm{(C)}}$, $ \lambda_\mathrm{S}$, $ \lambda_\mathrm{T}$, $\lambda_{\ell_1}$, $ \lambda_\mathrm{TSV}$, $ \lambda_{\tilde{X}})$ $=$ $ (10^{-3}$, $ 10^{-6}$, $ 10^{-5}$, $ 10^{-3}$, $ 10^{-3.5}$, $ 10^{-4})$. This selection was based on several criteria similar to those used in the test on the toy model, excluding comparisons with the ground truth.
In addition, we adopted the spatial correlation scale $\gamma=31^\circ$ and the temporal correlation scale $\tau=23\mathrm{[day]}$, following \citet{kawahara2020bayesian}, who estimated these values using Bayesian inference for cloud variability.

Figures~\ref{fig:dscovr_estimated_surface}, \ref{fig:dscovr_estimated_spectra}, and~\ref{fig:dscovr_estimated_cloud} display the estimated static surface distribution, the estimated spectra, and the snapshots of the dynamic component (Component 0) at three different dates.
For comparison, we show the reference cloud distribution in Figure~\ref{fig:dscovr_true_cloud}, which is the observed mean cloud fraction over 8 days provided by MODIS.
In the unmixed spectra, the strong oxygen B and A absorption features were observed at 0.688 and 0.764 \textmu m, respectively. Accordingly, Component 0 can be considered to have a flat spectrum, suggesting that it is cloud material. This is consistent with the assumption that Component 0 is a dynamic component.

For the surface distribution of Component 0, these snapshots capture some of the temporal features despite the limited spatial resolution. 
For instance, the rainy (March; left) and dry (August; middle) seasons in the Amazon (indicated by ``A''), the cloudy area in East Asia (indicated by ``B'' ) in August (middle), and the clear-sky area in North America (indicated by ``C'') in November (right) are captured.
In particular, the overall cloud pattern is best reproduced in March (left), likely because the most favorable viewing geometry is achieved at the equinox. 
Since the temporal resolution is largely determined by planetary rotation, patterns in the snapshot are elongated in the latitudinal direction. 
This elongation makes it impossible to retrieve the narrow and cloudy band at the equator known as the Intertropical Convergence Zone.
Thus, a valid reconstruction of the spatial distribution of Component 0 was achieved.

Moreover, we consider the static components. First, the spectrum of Component 1 reasonably reproduces the ocean spectrum, and the map also reflects the geography of the oceans. In particular, the improvements reported by \citet{kuwata2022global}, which appeared in the Atlantic and Indian Oceans, are still evident. Thus, we interpret that Component 1 corresponds to the oceans.
Regarding the spectra of Components 2 and 3, shapes similar to those in \citet[][static SOU with Tikhonov regularization]{kawahara2020global} were obtained. While the spectra estimated by \citet{kuwata2022global} were somewhat degenerate, the present result shows improved spectral resolution in the 0.5--0.6 \textmu m range. In practice, while Component 2 shows larger values than Component 3 at 0.688 and 0.764 \textmu m, it shows smaller values at 0.779 \textmu m; namely, Component 3 appears to have a redder spectrum than Component 2.
Furthermore, the increase at 0.688 and 0.764 \textmu m can be interpreted as Component 2 capturing the red edge of vegetation, although the strong oxygen absorption bands make the interpretation difficult. 
Therefore, we interpret that Component 2 corresponds to vegetation, and Component 3 corresponds to soil. 
Note that the interpretation of the retrieved components is based on knowledge of the Earth's surface spectra.
Although the spectral degeneracy has been somewhat resolved, the map still exhibits degeneracy, similar to \citet{kuwata2022global}.
Note that the North and South Poles are not represented very accurately due to the low observation weight for polar regions in DSCOVR \citep[][Figure 3(b)]{aizawa2020global}. While the results are consistent with the fact that Component 3 is observed in the Sahara Desert and around Chile, the degeneracy has not yet been completely resolved. This will be addressed in future work.



\begin{figure*}
  \centering
  {%
    \includegraphics[width=.95\textwidth]{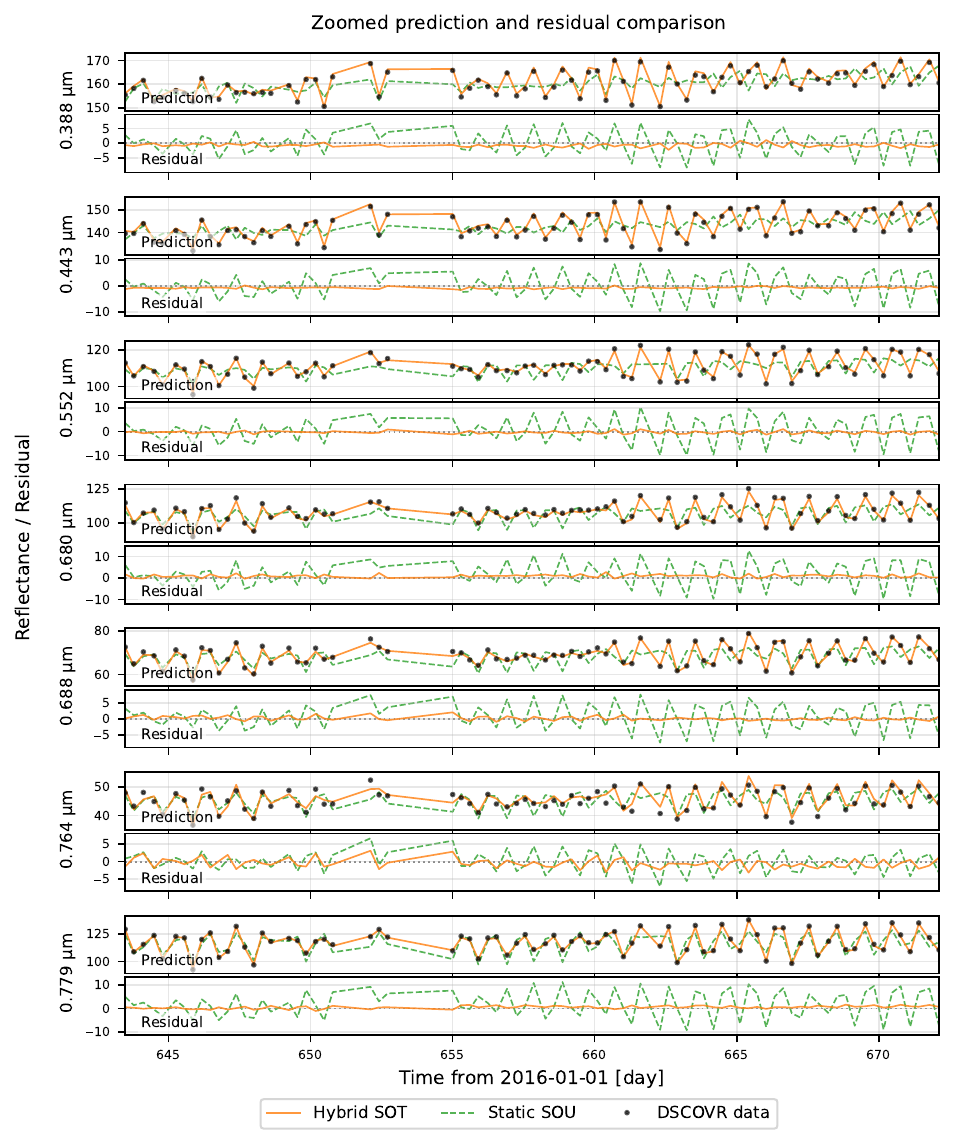}%
  } 
  \caption[]{
    Comparison of model retrieval performance between hybrid SOT and static SOU.
    Two panels are shown for each observational wavelength.
    The wavelengths are arranged from top to bottom as 0.388, 0.443, 0.552, 0.680, 0.688, 0.764, and 0.779 \textmu m.
    For each wavelength, the upper panel shows the observational data together with the model predictions.
    The observational data are shown in black.
    The model predictions from hybrid SOT and static SOU are shown in orange and green, respectively.
    The corresponding lower panel shows the residuals between the model predictions and the observational data.
  }
  \label{fig:zoom_prediction}
\end{figure*}

Next, we quantitatively evaluate the improvements achieved by hybrid SOT.
Figure~\ref{fig:zoom_prediction} shows a comparison of the observational data together with the model predictions from hybrid SOT and static SOU for each wavelength. 
The lower panels compare the residuals for the two methods, showing that hybrid SOT generally yields smaller residuals than static SOU.
For clarity, only a representative portion of the time series is shown here; the corresponding figure for the full time range is provided in Appendix~\ref{sec:appendix_lc_plot}.
Figure~\ref{fig:observation_vs_prediction} shows the relationship between the observational data and predictions for all wavelengths. The predictions from hybrid SOT are distributed more closely around the one-to-one line than those from static SOU, indicating better agreement with the observations.

\begin{figure}
  \centering
  {%
    \includegraphics[width=.48\textwidth]{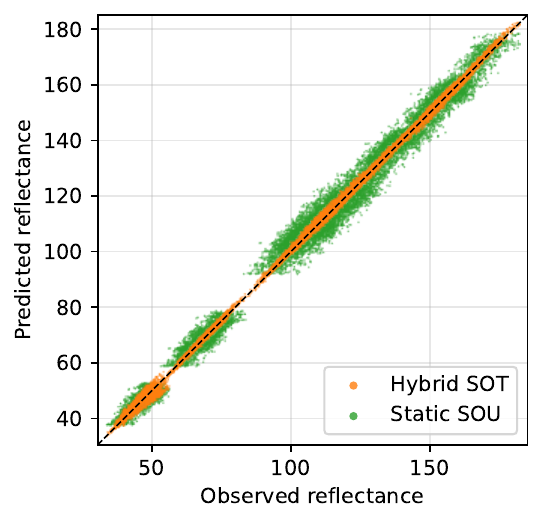}%
  } 
  \caption[]{Comparison of the observational data and predictions from hybrid SOT (orange) and static SOU (green) for all wavelengths. The diagonal dotted line indicates the one-to-one relation.}
  \label{fig:observation_vs_prediction}
\end{figure}

\begin{figure*}
  \centering
  {%
    \includegraphics[width=.99\textwidth]{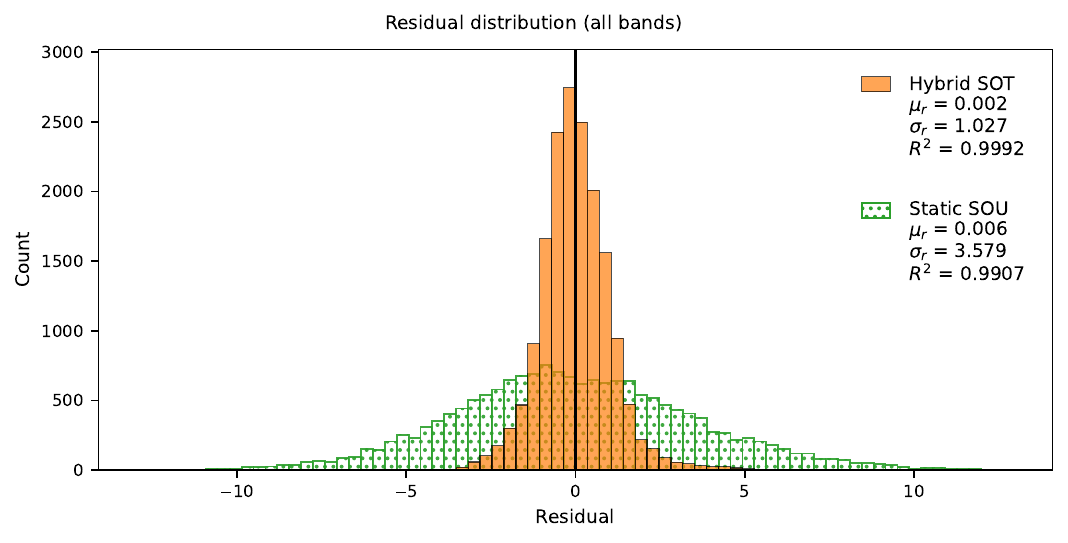}%
  }
  \caption[]{Histograms of the residuals between the real data and the predictions calculated using estimated solutions from hybrid SOT (orange) and static SOU (green).}
\label{fig:residual_histogram}
\end{figure*}

Furthermore, Figure~\ref{fig:residual_histogram} shows a histogram of all residuals.
If the model predictions perfectly matched the data, all residuals would be zero.
Thus, we can interpret that a narrower residual histogram indicates a better fit to the data.
In practice, the standard deviation of the residuals for hybrid SOT is $\sigma_r=1.027$, which is smaller than $\sigma_r=3.579$ for static SOU.
To evaluate this more quantitatively, we consider the coefficient of determination $R^2$, which measures how well a model reproduces the data. In typical applications, it takes values between 0 and 1, with a value of 1 indicating that the model prediction perfectly matches the data.
Let $\mu_r$  be the mean of the residuals. If $ |\mu_r| \ll \sigma_r$, we obtain
\begin{align}
    R^2 \simeq 1-\frac{\sigma_r^2}{\sigma_D^2},
\end{align}
where $\sigma_D$ denotes the standard deviation of the data. The details are provided in Appendix~\ref{sec:appendix_residual}. Thus, when $\sigma_r$ is small, corresponding to a narrow residual distribution, $R^2$ becomes larger. In fact, for the hybrid SOT, we obtained $R^2 = 0.9992$, confirming an improvement over the $R^2 = 0.9907$ observed for the static SOU.
Therefore, these results demonstrate that hybrid SOT can simultaneously reconstruct static and dynamic planetary components while achieving a significantly improved fit to real observational data.

%% file: revised_section_discussion.tex
\section{Discussion} \label{sec:discussion}

In our method, we employed $\ell_1$-norm and TSV regularization for the static surface distribution, Kronecker-sum kernel regularization and $\ell_1$-norm regularization for the dynamic geography, and volume regularization for the reflection spectra. 
Furthermore, we applied non-negativity constraints to all components; these originate specifically from NMF in spectral unmixing. 
These approaches allowed us to obtain solutions consistent with the intended effects of the regularization terms; however, alternative regularization schemes should also be investigated in future studies.
Several aspects can be modified, such as other types of sparse modeling, other regularization terms in spectral unmixing, kernels other than the RBF and Mat\'{e}rn-3/2 kernels, and forms other than the Kronecker sum.
It is necessary to verify how these estimates change when the regularization terms or constraints are modified. 
In particular, in our test, the degeneracy between the vegetation and soil components has not yet been sufficiently resolved, which is attributed to the indeterminacy of NMF, and future studies should investigate whether this degeneracy can be reduced through alternative regularization schemes.

Another important issue is the robustness of the proposed method under different observing conditions.
Although the toy model experiments demonstrate that hybrid SOT can simultaneously retrieve static and dynamic components across a range of observing geometries, a comprehensive robustness analysis covering a wider range of observing conditions remains an important subject for future work.
In practice, the retrieval performance depends not only on the observing conditions, such as the observational noise level, sampling rate $N_i$, and observing geometry, but also on the optimization settings, including the regularization parameters, initialization, and the optimization procedure.
Since the optimal choice of these parameters may depend on the observing conditions, a rigorous sensitivity analysis would require re-optimization for each configuration rather than fixing a single parameter set.
The experiments presented in Section~\ref{sec:test_toymap} and Appendix~\ref{app:geometry} provide a systematic assessment of the dependence of the retrieval results on observing geometry using a fixed set of hyperparameters. A broader investigation including other observing conditions and re-optimization for each configuration is beyond the scope of the present methodological study but will be essential for establishing practical observing requirements and evaluating the applicability of hybrid SOT to future observations.

Furthermore, for the forward model in this study, we used a simple sum of the static and dynamic components. However, considering the actual composition of the Earth, surface components in areas covered by clouds are obscured by the clouds, and the surface reflected light from those areas is not observed. 
Applying this fact, we could also consider a \textit{masking model} as the forward model:
\begin{align}
    D = \left( W \odot C \right)\bm{1}_{N_j} \bm{x}_\mathrm{C}^\top + \left(W \odot \left(\mathcal{I}-C\right)\right)AX, \label{eq:hybrid_masking_model}
\end{align}
where $\mathcal{I}$ denotes a matrix where all elements are one.
In this case, the constraint $0\le C_{ij} \le 1 $ $(i=1,\ldots, N_i, j=1,\ldots, N_j)$ would likely be necessary; however, optimization problems with additional bound constraints become somewhat more difficult to handle. 
Such an extension may be possible in the future.

The number of static and dynamic components is itself an important modeling assumption. 
Determining an appropriate model configuration, including the number of components and the choice of wavelength bands, remains an important topic for future studies.
Addressing this issue requires a more general forward model and systematic validation under various observing conditions.
In particular, extending hybrid SOT to multiple dynamic components would require a more flexible forward model capable of describing multiple time-dependent components, such as low- and high-altitude clouds.
The ultimate challenge is to make all components dynamic and to construct a Bayesian model. 
\citet{kawahara2020bayesian} introduced a Bayesian framework for single-component dynamic SOT.
In this study, we verified that integrating dynamic components is effective for inferring surface distribution and reflection spectra based on multi-color observations. 
Achieving multi-component dynamic global mapping with spectral unmixing would represent a significant milestone toward retrieving planetary components from photometric variability.

%% file: revised_section_conclusion.tex
\section{Conclusion} \label{sec:conclusion}

In this study, we proposed hybrid spin--orbit tomography, a method for the simultaneous retrieval of static and dynamic components, including their geographic distributions and reflection spectra. We integrated dynamic SOT \citep{kawahara2020bayesian} and static SOU \citep{kawahara2020global, kuwata2022global} to estimate a single dynamic component and $N_k$ static components.
We also verified that using a Kronecker-sum kernel rather than a Kronecker-product kernel as the regularization term for dynamic components (clouds) is effective. Overall, hybrid spin--orbit tomography provides a practical framework for simultaneously retrieving static and dynamic planetary components from unresolved multiband photometric observations. This work represents a step toward the characterization of directly imaged Earth-like exoplanets through future multiband photometric observations.

%% file: section_acknowledgments.tex
The authors are grateful to the DSCOVR team for making the data publicly available. We are indebted to Siteng Fan and Yuk L. Yung for providing the processed light curves and their geometric kernel from the DSCOVR dataset. We would also like to thank Kento Masuda, Masataka Aizawa, Shota Takahashi, Mirai Tanaka, and Shiro Ikeda for fruitful discussions. We would also like to thank the anonymous reviewer for the careful reading and constructive suggestions. This research was conducted using the Supermicro ARS-111GL-DNHR-LCC and FUJITSU Server PRIMERGY CX2550 M7 (Miyabi) at Joint Center for Advanced High Performance Computing (JCAHPC). This study was supported by JSPS KAKENHI grant Nos.\ 21H04998, 23H00133, 23H01224, 26H02072, 26K00752, and 26H02074 (H.K.). A.K.\ was also supported by JST SPRING, grant No.\ JPMJSP2108.

%% file: revised_section_kronecker_sum.tex
\section{Kronecker-Sum Kernel} \label{sec:kronecker}

\subsection{Regularization Term for the Kronecker-Product Kernel}

In this section, we derive the regularization term for the dynamic surface component $C \in \mathbb{R}^{N_i \times N_j}$ using a Bayesian framework.
Now, we define 
\begin{align}
    \bm{c} \coloneqq \mathrm{vec}(C), 
\end{align}
and the expanded weight matrix, which is an isomorphic form of $W$, as:
\begin{align}
    \tilde{W} \coloneqq\left( \diag\left(\hat{\bm{w}}_1\right)\ \cdots\ \diag\left(\hat{\bm{w}}_{N_j}\right)\right) \in\mathbb{R}^{N_i\times N_i N_j},
\end{align}
where $\diag(\cdot)$ is an operator to construct a diagonal matrix from a vector. Thus, we can rewrite Equation~\eqref{eq:dynamic_sot_Hadamard}, the forward model of dynamic SOT, as:
\begin{align}
    \bm{d} = \tilde{W} \bm{c} .
\end{align}

Here, the posterior distribution of $\bm{c}$ can be written as:
\begin{align}
    p\left( \bm{c} \middle| \bm{d} \right) = \frac{p\left( \bm{d} \middle| \bm{c} \right) p(\bm{c})}{p(\bm{d})},
\end{align}
and we assume that both the likelihood and the prior distribution follow Gaussian distributions:
\begin{align}
    p\left( \bm{d} \middle| \bm{c} \right) &= \mathcal{N}\left(\bm{d}\middle| \tilde{W} \bm{c} , \Sigma_{\bm{d}}\right), \\
    p(\bm{c}) &= \mathcal{N} \left(\bm{c} \middle| \bm{0}, \Sigma_{\bm{c}} \right),
\end{align}
where $\Sigma_{\bm{d}}$ and $\Sigma_{\bm{c}}$ denote the covariance matrices of the data and model, respectively. Then, we obtain:
\begin{align}
    p\left( \bm{c} \middle| \bm{d} \right) &\propto \exp \left( u(\bm{c}) \right), \label{eq:post_distribution}\\
    u(\bm{c}) &\coloneqq -\frac{1}{2}\left( \bm{d}-\tilde{W} \bm{c} \right)^\top \Sigma_{\bm{d}}^{-1} \left( \bm{d}-\tilde{W} \bm{c} \right) -\frac{1}{2} \bm{c}^\top \Sigma_{\bm{c}}^{-1} \bm{c}. \label{eq:post_distribution_u}
\end{align}
Let $\Sigma_{\bm{d}} = \sigma^2 I_{N_i}$ and $\Sigma_{\bm{c}} = \beta \left(K_\mathrm{S} \otimes K_\mathrm{T}\right)$, where $\sigma, \beta \in \mathbb{R}$, $K_\mathrm{S} \in \mathbb{R}^{N_j \times N_j}$, and $K_\mathrm{T}\in \mathbb{R}^{N_i \times N_i}$. Then, we can rewrite Equation~\eqref{eq:post_distribution_u} as:
\begin{align}
    u(\bm{c}) &= -\frac{1}{2\sigma^2}\left\| \bm{d}-\tilde{W} \bm{c} \right\|_2^2 -\frac{1}{2} \bm{c}^\top \left[\beta \left(K_\mathrm{S} \otimes K_\mathrm{T}\right)\right]^{-1} \bm{c} \\
    &= -\frac{1}{2\sigma^2}\left\| \bm{d}-\left(W \odot C\right) \bm{1}_{N_j} \right\|_2^2 -\frac{1}{2} \mathrm{vec}(C)^\top \left[\beta \left(K_\mathrm{S} \otimes K_\mathrm{T}\right)\right]^{-1} \mathrm{vec}(C). \label{eq:post_distribution_u_2}
\end{align}
Since maximum a posteriori (MAP) inference, which maximizes the posterior distribution, is equivalent to minimizing $-u(\bm{c})$, the optimization problem to solve can be written as:
\begin{align}
    \minimize_{C} \left\| \bm{d}-\left(W \odot C\right) \bm{1}_{N_j} \right\|_2^2 + \mathrm{vec}(C)^\top \left[\alpha \left(K_\mathrm{S} \otimes K_\mathrm{T}\right)\right]^{-1} \mathrm{vec}(C),
\end{align}
where $\alpha \coloneqq \beta / \sigma^2$.

In Section~\ref{sec:solve_optimization_problem}, we set the Kronecker-sum kernel, namely, $\Sigma_{\bm{c}} = \left(\lambda_\mathrm{S} K_{\mathrm{S}}^{-1} \oplus \lambda_\mathrm{T} K_{\mathrm{T}}^{-1}\right)^{-1}$. Note that in this case, the prior distribution is:
\begin{align}
    p(\bm{c}) &= \mathcal{N} \left(\bm{c} \middle| \bm{0}, \left(\lambda_\mathrm{S} K_{\mathrm{S}}^{-1} \oplus \lambda_\mathrm{T} K_{\mathrm{T}}^{-1}\right)^{-1} \right) \\
    & \propto \mathcal{N} \left(\bm{c} \middle| \bm{0}, \lambda_\mathrm{S}^{-1} K_{\mathrm{S}} \otimes I_{N_i} \right) \mathcal{N} \left(\bm{c} \middle| \bm{0},I_{N_j} \otimes \lambda_\mathrm{T}^{-1}
    K_{\mathrm{T}} \right),
\end{align}
which means that, using the Kronecker sum, the prior distribution can be interpreted as being proportional to the product of two probability distributions.


\subsection{Comparison between the Kronecker-Product and Kronecker-Sum Kernels}

\begin{figure*}
  \centering
  \subfigure[]{%
    \includegraphics[width=.99\textwidth]{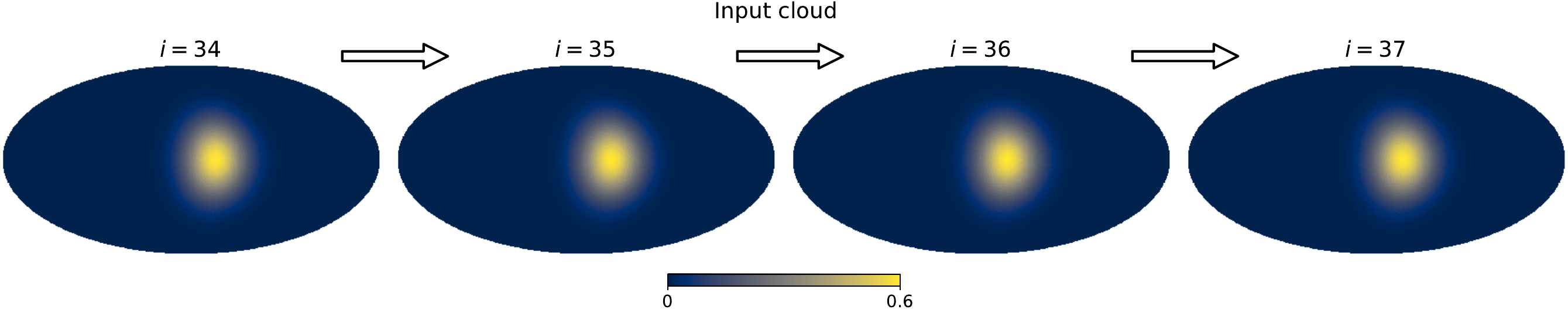}
    \label{fig:kronecker_true}%
  }
  \subfigure[]{%
    \includegraphics[width=.99\textwidth]{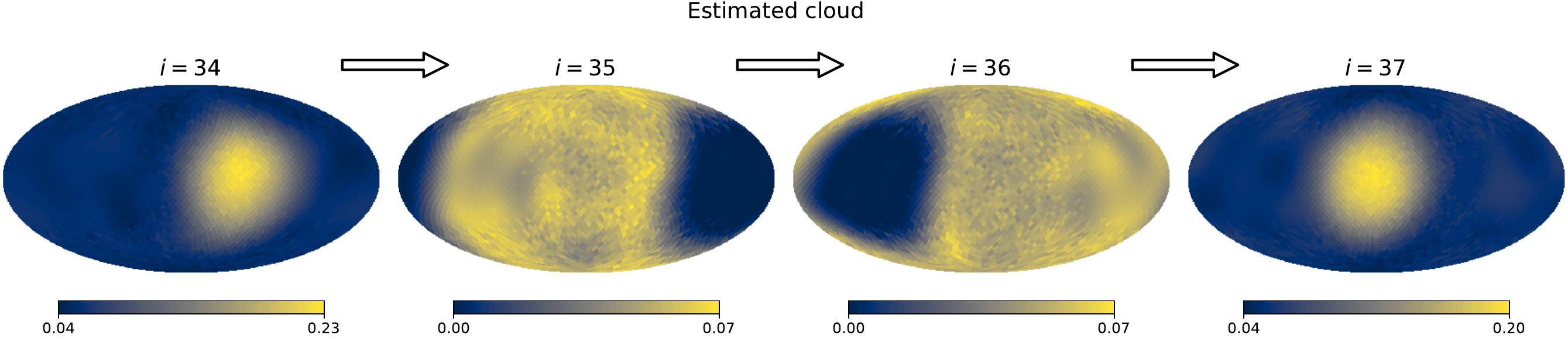}
    \label{fig:kronecker_estimated_product}%
  }
  \subfigure[]{%
    \includegraphics[width=.99\textwidth]{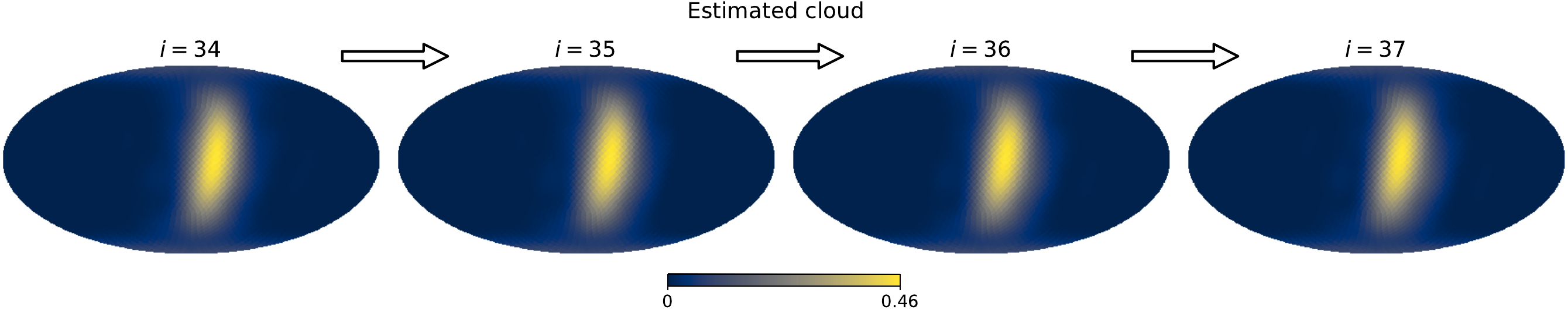}
    \label{fig:kronecker_estimated_sum}%
  }
  \caption[]{Comparison with Kronecker-product kernel and Kronecker-sum kernel. (a) True cloud distribution in the toy map. (b) The result of estimation using Kronecker-product kernel regularization. (c) The result of estimation using Kronecker-sum kernel regularization.}
\label{fig:kronecker_estimate}
\end{figure*}

In our method, we used a Kronecker-sum kernel as the regularization term. Here, we examine how the estimated solution changes when using a regularization term based on the Kronecker-product kernel or the Kronecker-sum kernel.
We use the toymap, including a dynamic component (cloud), which was generated in Section~\ref{sec:test_toymap}. Fixing the variables other than $C$, the reflection spectra, and static surface components, we estimate the dynamic surface distribution $C$. We employ a non-negativity constraint on $C$ and the following two types of regularization terms:
\begin{align}
    \alpha^{-1}\mathrm{vec}(C)^\top \left(K_{\mathrm{S}}^{-1} \otimes K_{\mathrm{T}}^{-1} \right) \mathrm{vec}(C), \\
    \mathrm{vec}(C)^\top \left(\lambda_\mathrm{S} K_{\mathrm{S}}^{-1} \oplus \lambda_\mathrm{T} K_{\mathrm{T}}^{-1}\right) \mathrm{vec}(C),
\end{align}
which are the same as Equation~\eqref{eq:RC_Kronecker_product} and \eqref{eq:RC_kronecker_sum}, respectively. For simplicity, we do not use $\ell_1$-norm regularization $(\lambda_{\ell_1}^{\mathrm{(C)}}=0)$.

Figure~\ref{fig:kronecker_estimate} shows the true cloud, the results using the Kronecker-product kernel, and the results using the Kronecker-sum kernel, with observation time frames $i=34,35,36,37$. We used the hyperparameters $\alpha^{-1} = 10^{-5}$, $\lambda_\mathrm{S}=10^{-5}$, and $\lambda_\mathrm{T}=10^{-5}$.
In the case of the Kronecker-product kernel, clouds appear to be located close to the true distribution at $i=34$ and $i=37$; however, when examining $i=35$ and $i=36$, it fails to capture the temporal variations of the clouds.
On the other hand, in the case of the Kronecker-sum kernel, clouds appear at the same positions as the true cloud, and the temporal variations are also captured. Thus, the Kronecker-sum kernel appears to be more effective in this optimization problem.
Note that the elongation of the clouds in the latitudinal direction is because the temporal resolution is largely determined by the spin rotation.

%% file: revised_section_appendix_proximal.tex
\section{Optimization Method} \label{sec:proximal}

As described in Section~\ref{sec:solve_optimization_problem}, we employ the proximal gradient method to solve the optimization problems. The proximal gradient method is one of the techniques used to solve optimization problems in which the objective function is expressed as the sum of differentiable and non-differentiable functions:
\begin{align}
    \minimize_{\bm{z}} f(\bm{z}) + \psi(\bm{z}),
\end{align}
where $f$ and $\psi$ denote a differentiable and non-differentiable function, respectively. The update equation is:
\begin{align}
    \bm{z}^{(i+1)} = \prox \left( \bm{z}^{(i)} -\gamma \nabla f \left(\bm{z}^{(i)}\right) \middle| \gamma \psi\right),
\end{align}
where $\prox$ is the proximal operator:
\begin{align}
    \prox (\bm{z} \mid \psi) \coloneqq \argmin_{\bm{w}} \left( \psi(\bm{w})+\frac{1}{2}\|\bm{w}-\bm{z}\|_2^2 \right).
\end{align}

Moreover, the optimization problem with non-negative constraints,
\begin{align}
    \minimize_{\bm{z}} f(\bm{z}) \subjectto \bm{z}\ge\bm{0},
\end{align}
is equivalent to the following optimization problem:
\begin{align}
    \minimize_{\bm{z}} f(\bm{z}) + \delta_+(\bm{z}),
\end{align}
where the non-differentiable function $\delta_+$ is defined as:
\begin{align}
     \delta_+(\bm{z}) \coloneqq 
        \begin{cases}  
            0 & ( \bm{z}\ge\bm{0} ) \\
            \infty & ( \mathrm{otherwise} ).
        \end{cases} \label{eq:nonnegative_indicator}
\end{align}
Therefore, the proximal gradient method can be applied as described above.

Note that the derivatives with respect to $C$ of the functions we used are as follows:
\begin{align}
    &\frac{\partial}{\partial C} F \left( C, \bm{x}_\mathrm{C}, A, X \right) \\
    &= \frac{\partial}{\partial C} \left\|D - \left( W \odot C \right)\bm{1}_{N_j} \bm{x}_C^\top - WAX \right\|_\mathrm{F}^2 \\
    &= -2 W \odot \left[ \left(D - \left( W \odot C \right)\bm{1}_{N_j} \bm{x}_C^\top - WAX\right) \bm{x}_C \bm{1}_{N_j}^\top \right], \\
    &\frac{\partial}{\partial C} \left[ \mathrm{vec}(C)^\top \left(\lambda_\mathrm{S} K_{\mathrm{S}}^{-1} \oplus \lambda_\mathrm{T} K_{\mathrm{T}}^{-1}\right) \mathrm{vec}(C) \right] \\
    &= 2 \lambda_\mathrm{S} C K_{\mathrm{S}}^{-1} + 2 \lambda_\mathrm{T} K_{\mathrm{T}}^{-1} C,
\end{align}
using the fact that $K_{\mathrm{S}}^{-1}$ and $K_{\mathrm{T}}^{-1}$ are symmetric. For derivatives with respect to other variables, see \citet[Section~3]{kuwata2022global}.

%% file: revised_section_evaluate.tex
\section{Choice of hyperparameters} \label{sec:evaluate}

We present a method for selecting the hyperparameters of the optimization problems. On the test using toymap (Section~\ref{sec:test_toymap}), we use four evaluation measures for reflection spectra and static surface components, as in \citet{kuwata2022global}:
(1) mean residual: the difference between the estimated model and observed data;
(2) $\det(\tilde{X}' \tilde{X}'^\top)$: the scaled volume regularization term;
(3) the scaled mean removed spectral angle ($\overline{\mathrm{MRSA}}$): the comparison of the estimated and true spectra; and
(4) the correct pixel rate (CPR): the comparison of the estimated and true static map.
The definition and details of these measures are explained in \citet[Appendix~E]{kuwata2022global}.

In addition, we also use two evaluation measures for the dynamic surface components. The first is the root mean square error (RMSE):
\begin{align}
    \mathrm{RMSE} \coloneqq \sqrt{\frac{\left\| C^{*} - C_\mathrm{true} \right\|_\mathrm{F}^2 }{N_i N_j}},
\end{align}
where $C^{*}$ and $ C_\mathrm{true}$ are the estimated and true distributions, respectively. RMSE allows for a direct comparison of these.
The second is the correlation coefficient:
\begin{align}
    \mathrm{corr}\left(C^{*}, C_\mathrm{true} \right) \coloneqq
    \frac{ \left(\bm{c}^{*} - \overline{c^{*}} \bm{1}_{N_i N_j}\right)^\top \left(\bm{c}_\mathrm{true} - \overline{c_\mathrm{true}} \bm{1}_{N_i N_j}\right) }
    { \left\|\bm{c}^{*} - \overline{c^{*}} \bm{1}_{N_i N_j}\right\|_2 \left\|\bm{c}_\mathrm{true} - \overline{c_\mathrm{true}} \bm{1}_{N_i N_j}\right\|_2 },
\end{align}
where $\bm{c}^{*} \coloneqq \mathrm{vec}(C^{*})$; $\bm{c}_\mathrm{true} \coloneqq \mathrm{vec}(C_\mathrm{true})$; and $\overline{c^{*}}$ and $\overline{c_\mathrm{true}}$ denote the average of $\bm{c}^{*}$ and $\bm{c}_\mathrm{true}$, respectively.

\begin{figure*}
  \centering
  \subfigure[]{%
    \includegraphics[width=.32\textwidth]{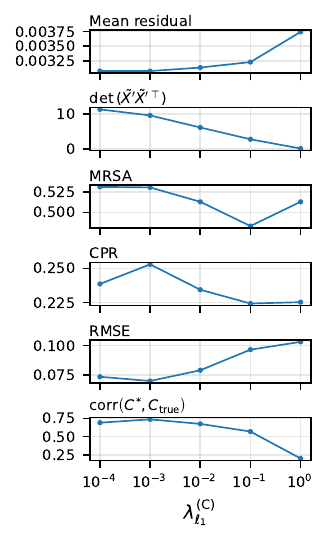}%
  }
  \subfigure[]{%
    \includegraphics[width=.32\textwidth]{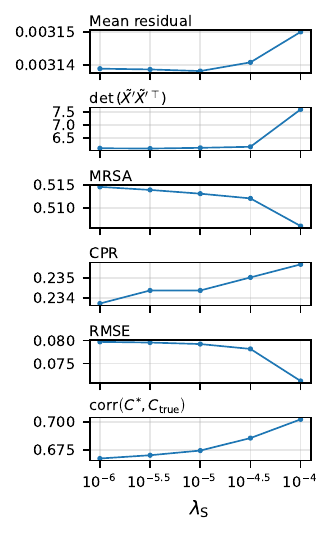}%
  }
  \subfigure[]{%
    \includegraphics[width=.32\textwidth]{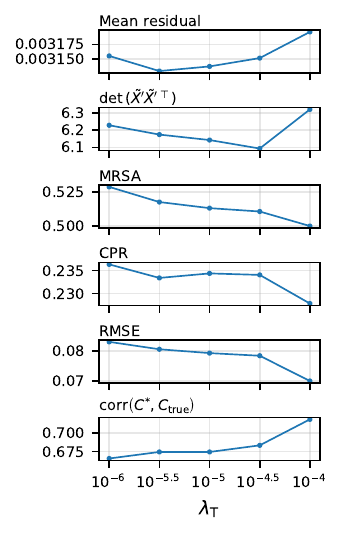}%
  }
  \subfigure[]{%
    \includegraphics[width=.32\textwidth]{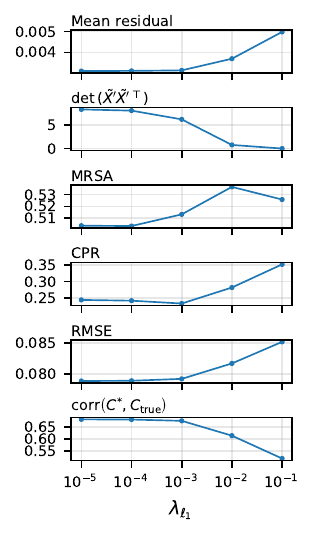}%
  }
  \subfigure[]{%
    \includegraphics[width=.32\textwidth]{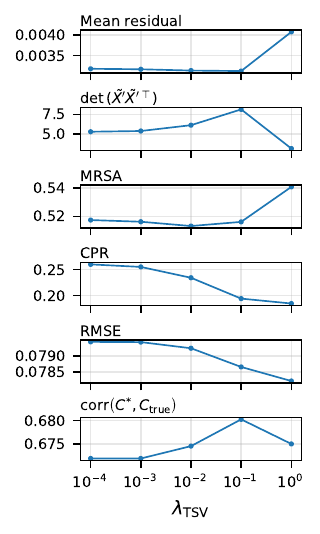}%
  }
  \subfigure[]{%
    \includegraphics[width=.32\textwidth]{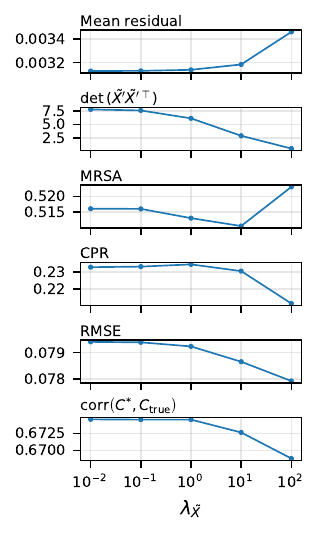}%
  }
  \caption[]{Evaluation measures for the test using toymap.
  The values are calculated by fixing the parameters and varying only one parameter, 
  (a) $ \lambda_{\ell_1}^{\mathrm{(C)}}$, 
  (b) $ \lambda_\mathrm{S}$, 
  (c) $ \lambda_\mathrm{T}$, 
  (d) $ \lambda_{\ell_1}$, 
  (e) $ \lambda_\mathrm{TSV}$, 
  and (f) $ \lambda_{\tilde{X}} $.}
\label{fig:evaluate_toymap}
\end{figure*}

Figure~\ref{fig:evaluate_toymap} shows the values of each measure using estimated solutions. To select hyperparameters, we calculate these values by moving one parameter and fixing the other parameters. For example, when we select
$ \lambda_{\ell_1}^{\mathrm{(C)}}$, we calculate the measures by varying $ \lambda_{\ell_1}^{\mathrm{(C)}}$ while fixing $ \lambda_\mathrm{S}$, $ \lambda_\mathrm{T}$, $ \lambda_{\ell_1}$, $ \lambda_\mathrm{TSV}$, $ \lambda_{\tilde{X}} $. 
By definition, we should select regularization parameters that resulted in smaller mean residuals, $\det(\tilde{X}' \tilde{X}'^\top)$, $\overline{\mathrm{MRSA}}$, and RMSE, and larger CPR and $\mathrm{corr}\left(C^{*}, C_\mathrm{true} \right)$.
There tends to be a trade-off between the mean residual and the scaled volume regularization term. Thus, we selected the hyperparameters to avoid large values for the mean residual and the scaled volume regularization term, while also referring to $\overline{\mathrm{MRSA}}$, CPR, RMSE, and $\mathrm{corr}\left(C^{*}, C_\mathrm{true} \right)$. 
Consequently, on the test using toymap, we adopted ( $ \lambda_{\ell_1}^{\mathrm{(C)}}$, $ \lambda_\mathrm{S}$, $ \lambda_\mathrm{T}$, $ \lambda_{\ell_1}$, $ \lambda_\mathrm{TSV}$, $ \lambda_{\tilde{X}} $ ) $=$ ( $ 10^{-2}$, $ 10^{-5}$, $ 10^{-5}$, $ 10^{-3}$, $ 10^{-2}$, $ 10^{0}$ ).

\begin{figure*}
  \centering
  \subfigure[]{%
    \includegraphics[width=.32\textwidth]{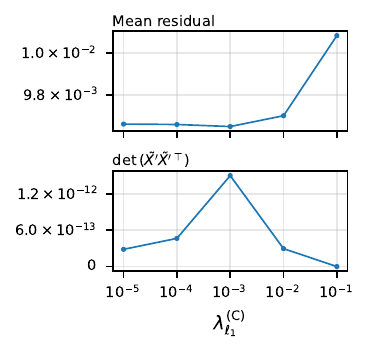}%
  }
  \subfigure[]{%
    \includegraphics[width=.32\textwidth]{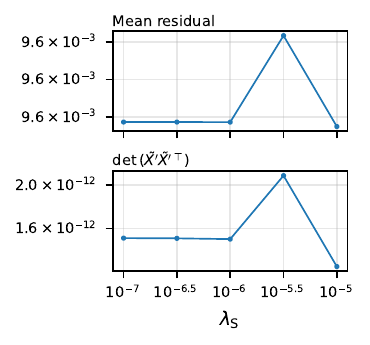}%
  }
  \subfigure[]{%
    \includegraphics[width=.32\textwidth]{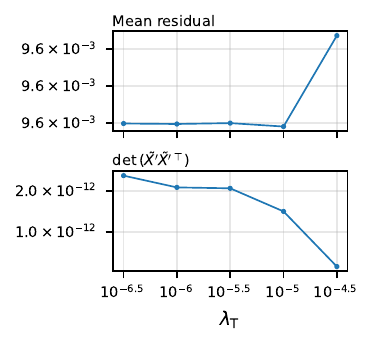}%
  }
  \subfigure[]{%
    \includegraphics[width=.32\textwidth]{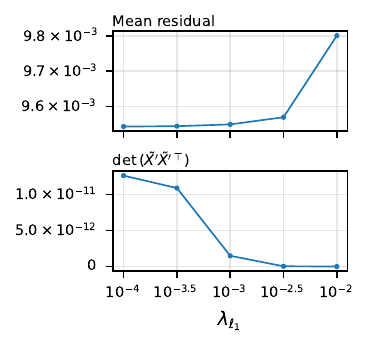}%
  }
  \subfigure[]{%
    \includegraphics[width=.32\textwidth]{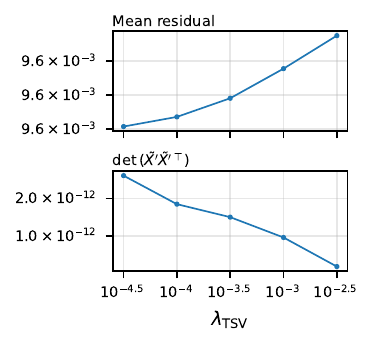}%
  }
  \subfigure[]{%
    \includegraphics[width=.32\textwidth]{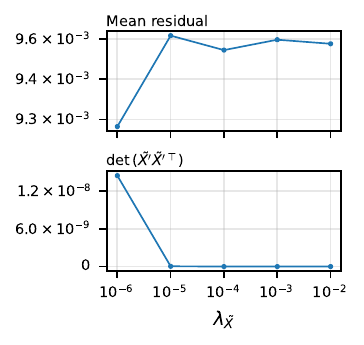}%
  }
  \caption[]{Evaluation measures for the test using DSCOVR data.
  The values are calculated by fixing the parameters and varying only one parameter, 
  (a) $ \lambda_{\ell_1}^{\mathrm{(C)}}$, 
  (b) $ \lambda_\mathrm{S}$, 
  (c) $ \lambda_\mathrm{T}$, 
  (d) $ \lambda_{\ell_1}$, 
  (e) $ \lambda_\mathrm{TSV}$, 
  and (f) $ \lambda_{\tilde{X}} $.}
\label{fig:evaluate_dscovr}
\end{figure*}

On the test using DSCOVR data (Section~\ref{sec:test_dscovr}), we cannot use the scaled $\overline{\mathrm{MRSA}}$, RMSE, CPR, or $\mathrm{corr}\left(C^{*}, C_\mathrm{true} \right)$ since it is not possible to calculate these values for actual exoplanet observations because the ground truth is unknown.
Thus, we selected the regularization parameters by examining the mean residual and $\det(\tilde{X}' \tilde{X}'^\top)$ while checking the geographical features of the map and the shape of the spectra. 
Figure~\ref{fig:evaluate_dscovr} shows the values of each measure using estimated solutions on the test using DSCOVR data. 
Consequently, we adopted ( $ \lambda_{\ell_1}^{\mathrm{(C)}}$, $ \lambda_\mathrm{S}$, $ \lambda_\mathrm{T}$, $\lambda_{\ell_1}$, $ \lambda_\mathrm{TSV}$, $ \lambda_{\tilde{X}} $ ) $=$ ( $ 10^{-3}$, $ 10^{-6}$, $ 10^{-5}$, $ 10^{-3}$, $ 10^{-3.5}$, $ 10^{-4}$ ).

%% file: section_appendix_geometry.tex
\section{Sensitivity to Observing Geometry}
\label{app:geometry}

In Section~\ref{sec:test_toymap}, we demonstrated hybrid SOT using a toy Earth model.
In this section, we further examine the effects of observing geometry on the retrieval results.

\begin{figure*}
\centering
{%
\includegraphics[width=.95\textwidth]{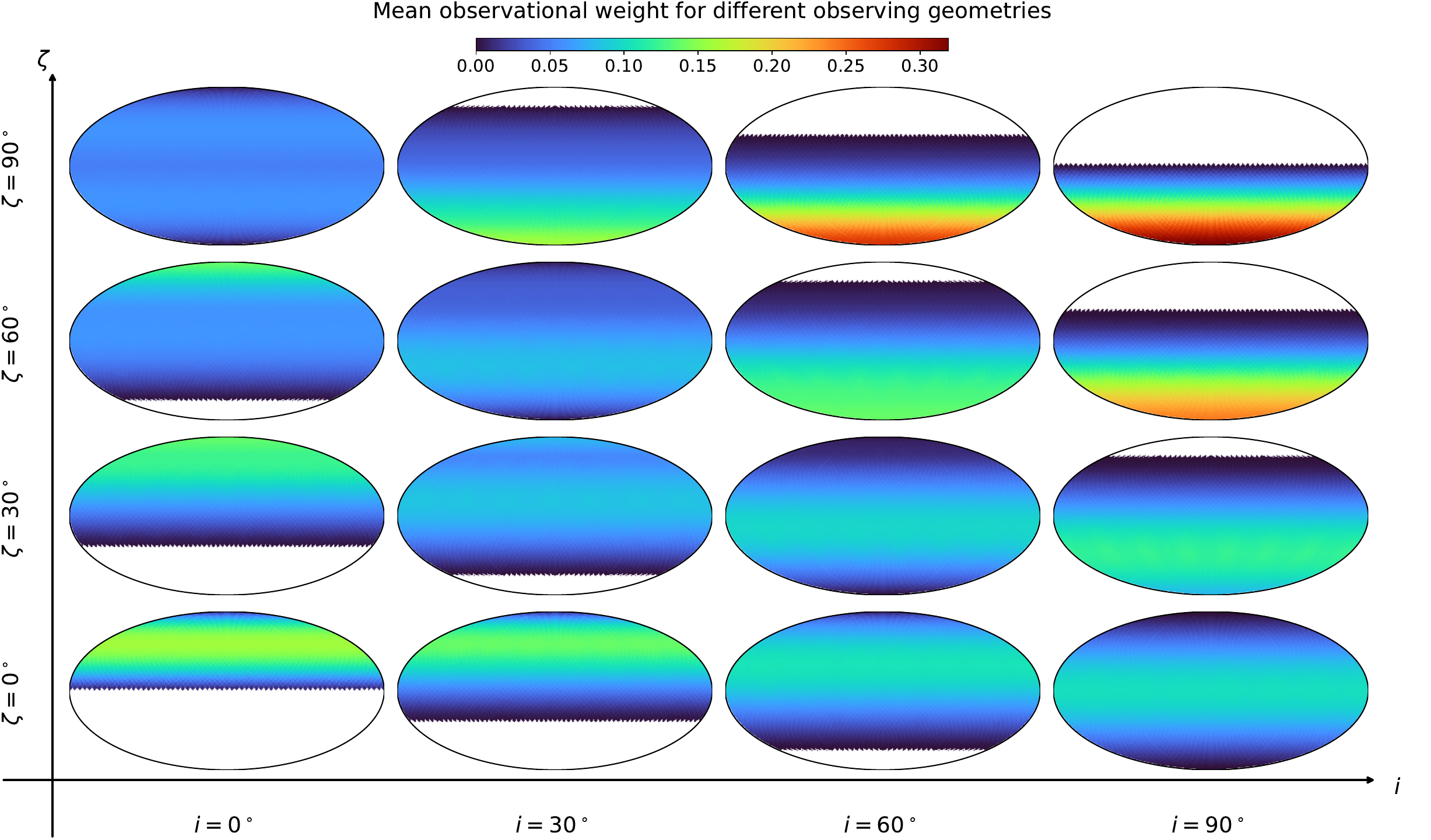}%
}
\caption[]{Time-averaged observational weight for $4 \times 4$ different observing geometries. Areas where the value is zero are shown in white.}
\label{fig:geometry_weight}
\end{figure*}

We constructed toy models for $4 \times 4$ combinations of the orbital inclination and obliquity, with $i = 0^\circ, 30^\circ, 60^\circ,$ and $90^\circ$ and
$\zeta = 0^\circ, 30^\circ, 60^\circ,$ and $90^\circ$.
Here, the quantity that depends on the observing geometry is $W$, which is determined by the illuminated and visible area of the surface.
Figure~\ref{fig:geometry_weight} shows the time-averaged observational weight,
\begin{align}
\overline{W}_{j} = \frac{1}{N_i}\sum_{i} W_{ij},
\end{align}
for each geometry.
Areas where $\overline{W}_{j}=0$ are shown in white; these areas were never observed during the observation period.
For instance, when $(i, \zeta) = (0^\circ, 0^\circ)$ (the lower-left panel), the northern hemisphere contributes to the disk-integrated observations, but the southern hemisphere does not.

\begin{figure*}
\centering
{%
\includegraphics[width=.95\textwidth]{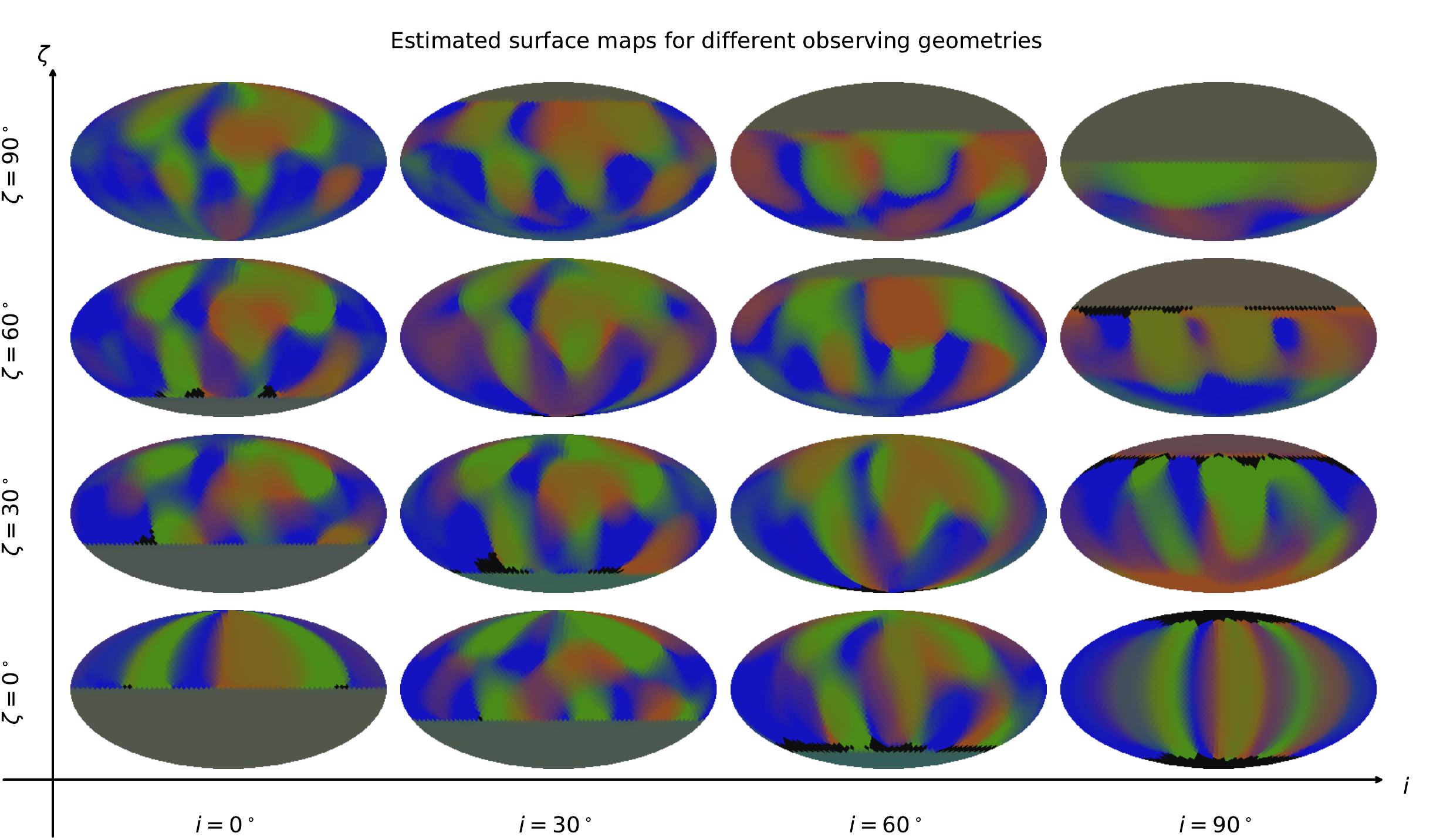}%
}
\caption[]{Color composite of the estimated static surface distribution for $4 \times 4$ different observing geometries. Pixels where all components are zero are shown in black.}
\label{fig:geometry_surface}
\end{figure*}

We applied the proposed method to the toy models for the $4 \times 4$ observing geometries using the same set of hyperparameters for all configurations:
$(\lambda_{\ell_1}^{\mathrm{(C)}}$, $ \lambda_\mathrm{S}$, $ \lambda_\mathrm{T}$, $ \lambda_{\ell_1}$, $ \lambda_\mathrm{TSV}$, $ \lambda_{\tilde{X}})$ $=$ $(10^{-2}$, $ 10^{-5}$, $ 10^{-5}$, $ 10^{-3}$, $ 10^{-2}$, $ 10^{0})$ .
These values are the same as those adopted for the fiducial toy-model experiment in Section~\ref{sec:test_toymap}; fixing them across all configurations allows us to isolate the effect of the observing geometry.
Figure~\ref{fig:geometry_surface} shows the estimated static surface distributions for each observing geometry.
Although no explicit constraints or regularizations based on the observational weights were employed, the retrieved component amplitudes are negligible in areas where the observational weights are zero.
In practice, the southern hemisphere in the $(i, \zeta) = (0^\circ, 0^\circ)$ panel and the northern hemisphere in the $(i, \zeta) = (90^\circ, 90^\circ)$ panel have nearly zero values across all components and are depicted with a color that appears to be a mixture of all colors.
Note that pixels where all components are zero are rendered in black; this likely results from the sparsity-promoting optimization in regions that are weakly constrained by the data.
Moreover, in $(i, \zeta) = (60^\circ, 90^\circ)$ and $(90^\circ, 60^\circ)$, the low retrieval accuracy in the low- to mid-latitude regions may be associated with the concentration of observational weight around Antarctica.
In contrast, the other panels generally show better retrieval of continental shapes and separation of the surface components.

\begin{figure*}
\centering
\subfigure[]{%
\includegraphics[width=.99\textwidth]{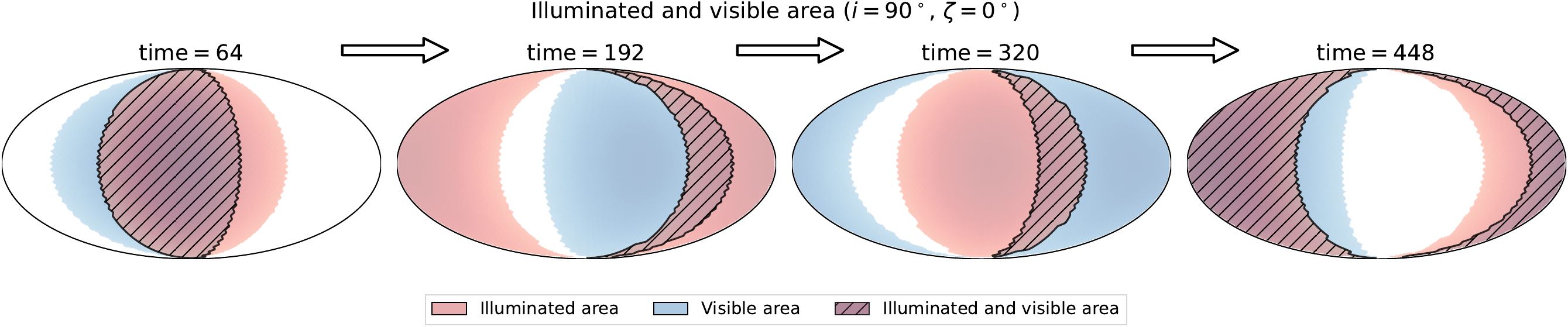}
\label{fig:IVarea_i90_zeta0}%
}
\subfigure[]{%
\includegraphics[width=.99\textwidth]{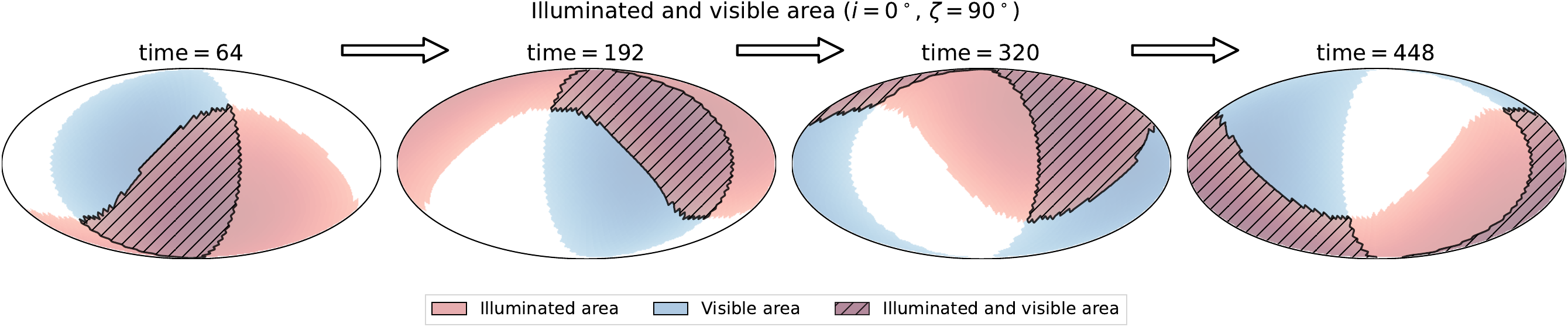}
\label{fig:IVarea_i0_zeta90}%
}
\caption[]{Temporal changes in the illuminated and visible area of the planetary surface in (a) $(i, \zeta) =(90^\circ, 0^\circ)$ and (b) $(i, \zeta) = (0^\circ, 90^\circ)$. Red indicates the illuminated area, and blue indicates the visible area. Their intersection, corresponding to the illuminated and visible area, is hatched with diagonal lines.}
\label{fig:IVarea}
\end{figure*}

Here, we focus on the panels in  $(i, \zeta) = (0^\circ, 90^\circ)$ and $(90^\circ, 0^\circ)$.
Although the time-averaged observational weight extends over nearly the entire surface in both configurations, the estimated maps differ substantially.
While the map at $(i, \zeta) = (0^\circ, 90^\circ)$ is well retrieved, the map at $(i, \zeta) = (90^\circ, 0^\circ)$ shows continental structures that are symmetrically elongated about the equator in the latitudinal direction.
To understand this, Figure~\ref{fig:IVarea} shows the temporal evolution of the illuminated and visible area of the surface, which determines the distribution of $W$ over the surface (see Section~\ref{sec:SOT}).
In the case of $(i, \zeta) = (90^\circ, 0^\circ)$ (Figure~\ref{fig:IVarea_i90_zeta0}), the illuminated and visible areas are symmetric about the equator at all times.
Then, the observational kernel cannot distinguish contributions from surface elements at mirrored northern and southern latitudes, producing a north--south degeneracy in the data.
Consequently, the resolution in the latitudinal direction is reduced, which may lead to the latitudinal elongation seen in the estimated map.
On the other hand, at $(i, \zeta) = (0^\circ, 90^\circ)$, the illuminated and visible area moves across different parts of the surface during the observation period. Therefore, it can be inferred that the ability to retrieve a two-dimensional map by solving the inverse problem arises from the two-dimensional spatial information contained in the observational data.

\begin{figure*}
\centering
\subfigure[]{%
\includegraphics[width=.99\textwidth]{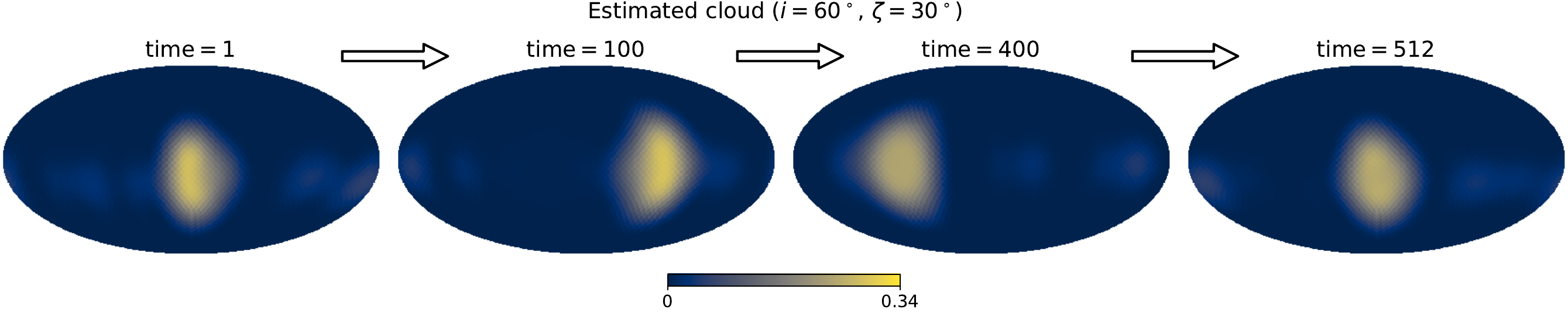}
\label{fig:cloud_i60_zeta30}%
}
\subfigure[]{%
\includegraphics[width=.99\textwidth]{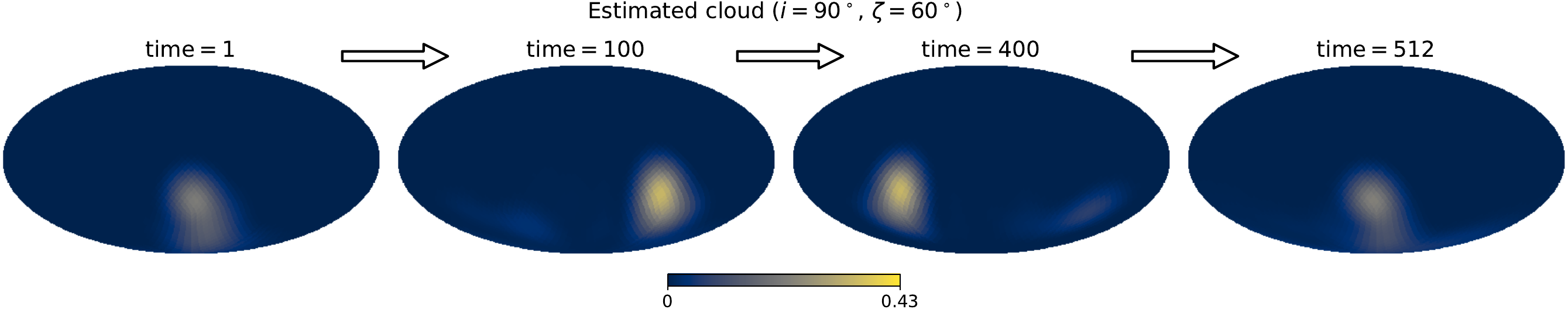}
\label{fig:cloud_i90_zeta60}%
}
\subfigure[]{%
\includegraphics[width=.99\textwidth]{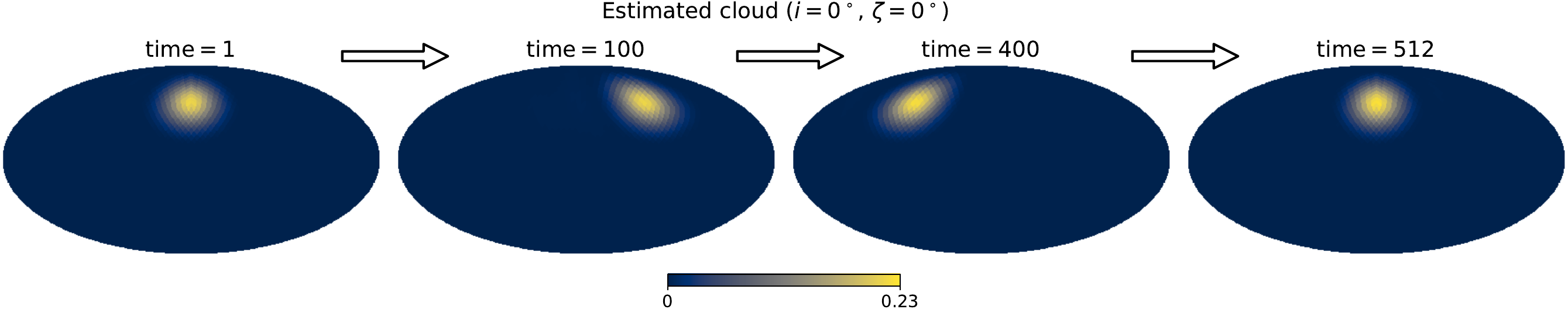}
\label{fig:cloud_i0_zeta0}%
}
\caption[]{Estimated dynamic component (cloud) distributions in (a) $(i, \zeta) = (60^\circ, 30^\circ)$, (b) $(i, \zeta) = (90^\circ, 60^\circ)$, and (c) $(i, \zeta) = (0^\circ, 0^\circ)$.}
\label{fig:geometry_cloud}
\end{figure*}

Figure~\ref{fig:geometry_cloud} shows representative examples of the estimated dynamic cloud distribution.
The longitudinal motion of the clouds could be retrieved across all configurations, as shown by the $(i, \zeta) = (60^\circ, 30^\circ)$ case (Figure~\ref{fig:cloud_i60_zeta30}).
Although the estimated clouds appear more spatially extended than the input cloud (Figure~\ref{fig:init_cloud}), this may be reduced by tuning the cloud-related hyperparameters.
However, as seen for $(i, \zeta) = (90^\circ, 60^\circ)$ and $(0^\circ, 0^\circ)$  (Figure~\ref{fig:cloud_i90_zeta60} and \ref{fig:cloud_i0_zeta0}, respectively), the estimated cloud positions can be affected in configurations with strongly nonuniform observational weights.

\begin{figure*}
\centering
{%
\includegraphics[width=.99\textwidth]{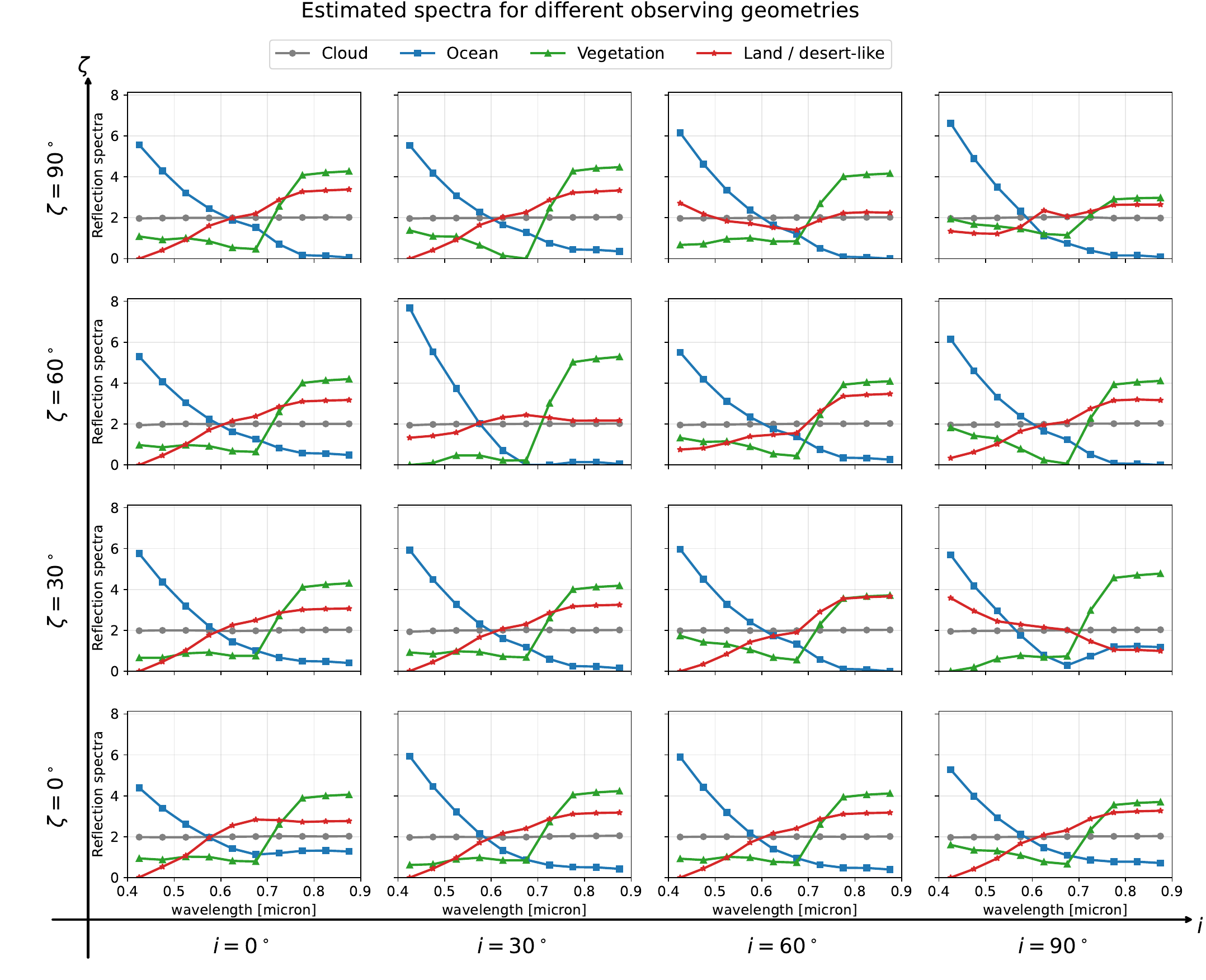}%
}
\caption[]{Estimated static and dynamic spectra for $4 \times 4$ different observing geometries.}
\label{fig:geometry_spectra}
\end{figure*}

Furthermore, Figure~\ref{fig:geometry_spectra} shows the estimated reflection spectra for each observing geometry.
The shapes of the dynamic and static components are successfully retrieved in many configurations.
In $(i, \zeta) = (30^\circ, 60^\circ)$ and $(90^\circ, 60^\circ)$, the spectrum of the land appears to be partially degenerate with those of other components; however, this may be improved through hyperparameter tuning.
Similar spectral degeneracies are seen for $(i, \zeta) = (60^\circ, 60^\circ)$ and $(90^\circ, 90^\circ)$, although they may also be associated with the distribution of the observational weights.

Overall, these experiments demonstrate that hybrid SOT can retrieve static and dynamic spatial distributions, together with their spectra, across a wide range of observing geometries.
However, retrieval accuracy can degrade when the observational weights are strongly nonuniform or when the observing geometry introduces spatial degeneracies.
Since the same hyperparameters are used for all configurations in this test, these results characterize the geometry dependence of a fixed retrieval setup.
A more quantitative assessment of the achievable performance for individual observing geometries would require re-optimization of the hyperparameters for each configuration and is left for future work.

%% file: section_appendix_residual.tex
\section{Evaluation of DSCOVR Estimation} \label{sec:appendix_evaluation_dscovr}

\subsection{Visual Evaluation} \label{sec:appendix_lc_plot}

\begin{figure*}
  \centering
  {%
    \includegraphics[width=.99\textwidth]{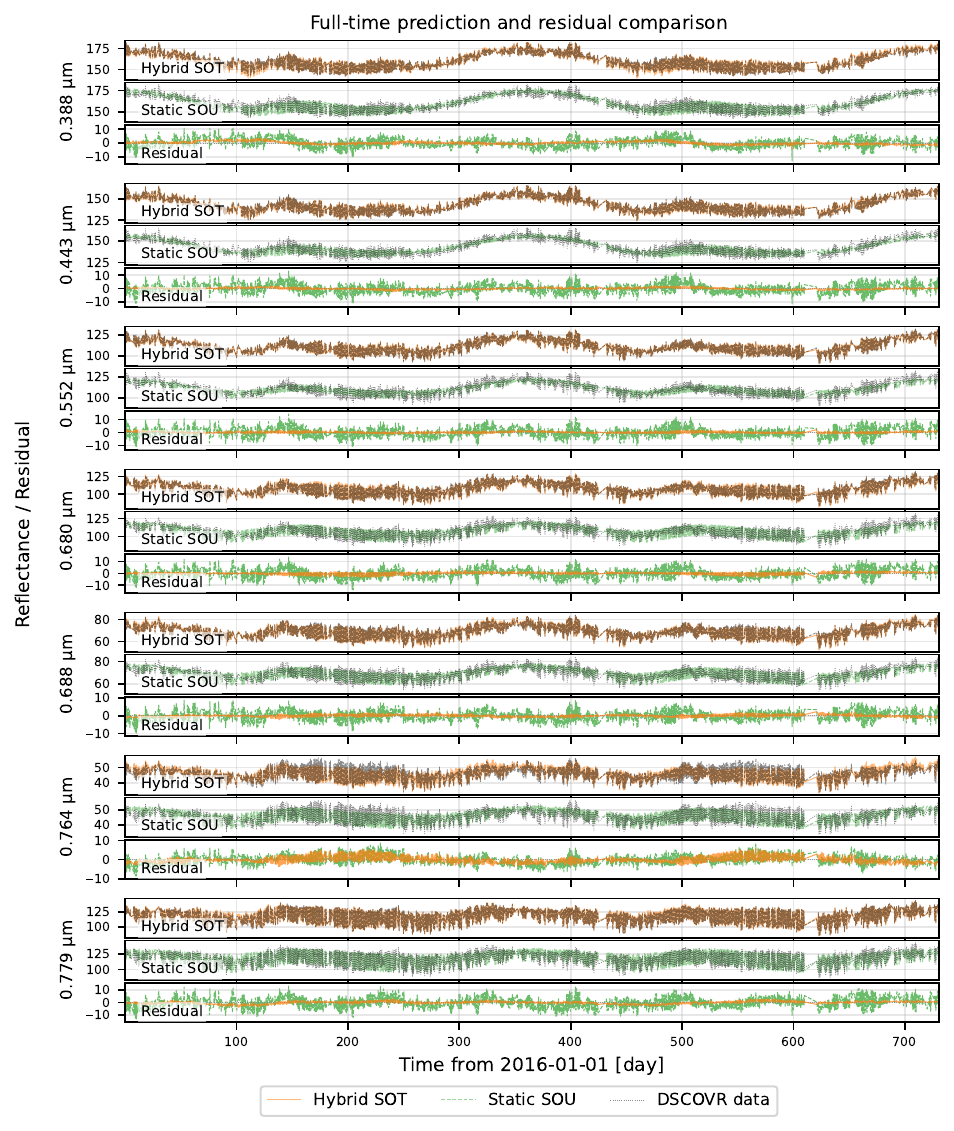}%
  }
  \caption[]{For each wavelength, the real observational data together with the model predictions from hybrid SOT (top) and static SOU (middle), and the residuals between the model predictions and the observational data (bottom). The residuals for hybrid SOT and static SOU are shown in orange and green, respectively.}
\label{fig:full_prediction}
\end{figure*}

Section~\ref{sec:test_dscovr} presents a comparison between the observational data, the model predictions from hybrid SOT and static SOU, and their corresponding residuals using a representative portion of the time series.
Figure~\ref{fig:full_prediction} shows the corresponding comparison over the full observation time range.
These results confirm that hybrid SOT captures temporal variations that static SOU does not.

\subsection{Quantitative Evaluation: Coefficient of Determination} \label{sec:appendix_residual}

In Section~\ref{sec:test_dscovr}, the \textit{coefficient of determination} $R^2$ was used to evaluate the estimated solutions obtained from DSCOVR:
\begin{align}
    R^2 \coloneqq 1- \frac{ \left\| D-D^* \right\|_\mathrm{F}^2}{\left\| D- \mu_D \mathcal{I} \right\|_\mathrm{F}^2}, \label{eq:coefficient_of_determination}
\end{align}
where $D^*$ denotes the prediction obtained using the estimated solutions $C^*$, $\bm{x}_\mathrm{C}^*$, $A^*$, and $X^*$; and $\mu_D$ denotes the mean of the real data $D$.
The coefficient of determination measures how well a model reproduces the data, which takes a value of 1 when the model prediction perfectly matches the real data, while it takes a value of 0 when the model predicts only the mean. Note that $R^2$ may take on a negative value if the prediction accuracy is lower than that of the mean.

First, we define the residual $r_{il} \coloneqq D_{il} - D_{il}^*$ $(i=1,\ldots,N_i,l=1,\ldots,N_l)$. 

The mean and the standard deviation of $r_{il}$ are
\begin{align}
    \mu_r &= \frac{1}{N_i N_l} \sum_i \sum_l r_{il}, \\
    \sigma_r &= \sqrt{ \frac{1}{N_i N_l} \sum_i \sum_l (r_{il} - \mu_r )^2 }.
\end{align}
Moreover, the standard deviation of the data is
\begin{align}
    \sigma_D = \sqrt{ \frac{1}{N_i N_l} \sum_i \sum_l (D_{il} - \mu_D )^2 }.
\end{align}
Using these quantities, we can rewrite Equation~\eqref{eq:coefficient_of_determination} as
\begin{align}
    R^2 = 1 - \frac{\sigma_r^2 + \mu_r^2}{\sigma_D^2}.
\end{align}
If $ |\mu_r| \ll \sigma_r$, then $\sigma_r^2 + \mu_r^2 \simeq \sigma_r^2$. Thus, we obtain
\begin{align}
    R^2 \simeq 1-\frac{\sigma_r^2}{\sigma_D^2}.
\end{align}
Therefore, a smaller $\sigma_r$, corresponding to a narrower residual distribution, results in a larger $R^2$.